\documentclass[aps,prd,reprint,superscriptaddress,nofootinbib]{revtex4-1}
\usepackage{graphicx}
\usepackage{subfig}
\usepackage{multirow} 
\usepackage{amssymb}
\usepackage{amsmath}
\usepackage{enumerate}
\usepackage{todonotes}
\usepackage[normalem]{ulem}
\usepackage{ragged2e}

\newcommand{\ZHAireS}{\mbox{ZHA\scriptsize{${\textrm{IRE}}$}\normalsize{\hspace{.05em}S}}\hspace{.45em}}

\def \nutau {$\nu_\tau~$}

\newcommand{\atUC}{\affiliation{Dept. of Physics, Enrico Fermi Inst., Kavli Inst. for Cosmological Physics, Univ. of Chicago, Chicago, IL 60637 USA.}}
\newcommand{\atUCLA}{\affiliation{Dept. of Physics and Astronomy, Univ. of California, Los Angeles, Los Angeles, CA 90095 USA.}}
\newcommand{\atOSU}{\affiliation{Dept. of Physics, Center for Cosmology and AstroParticle Physics, Ohio State Univ., Columbus, OH 43210 USA.}}
\newcommand{\atUH}{\affiliation{Dept. of Physics and Astronomy, Univ. of Hawaii, Manoa, HI 96822.}}
\newcommand{\atNTU}{\affiliation{Dept. of Physics, Grad. Inst. of Astrophys.,\& Leung Center for Cosmology and Particle Astrophysics, National Taiwan University, Taipei, Taiwan.}}
\newcommand{\atKU}{\affiliation{Dept. of Physics and Astronomy, Univ. of Kansas, Lawrence, KS 66045 USA.}}
\newcommand{\atWU}{\affiliation{Dept. of Physics, McDonnell Center for the Space Sciences, Washington Univ. in St. Louis, MO 63130 USA.}}
\newcommand{\atSLAC}{\affiliation{SLAC National Accelerator Laboratory, Menlo Park, CA, 94025 USA.}}
\newcommand{\atUD}{\affiliation{Dept. of Physics, Univ. of Delaware, Newark, DE 19716 USA.}}
\newcommand{\atUCL}{\affiliation{Dept. of Physics and Astronomy, University College London, London, United Kingdom.}}
\newcommand{\atJPL}{\affiliation{Jet Propulsion Laboratory, California Institute of Technology, Pasadena, CA 91109 USA.}}

\newcommand{\atBRUS}{\affiliation{Astrophysical Institute, Vrije Universiteit Brussel, Pleinlaan 2, 1050, Brussels, Belgium.}}
\newcommand{\atWi}{\affiliation{Department of Physics, University of Wisconsin-Madison, Madison, WI 53706 USA.}}
\newcommand{\atUCSD}{\affiliation{Center for Astrophysics and Space Sciences, Univ. of California, San Diego, La Jolla, CA 92093 USA.}}
\newcommand{\atMoscow}{\affiliation{Moscow Engineering Physics Institute, Moscow, Russia.}}

\begin{document}


\title{
A comprehensive analysis of anomalous ANITA events disfavors a diffuse tau-neutrino flux origin
}
\author{A. Romero-Wolf}\atJPL
\author{S.~A.~Wissel} 
\affiliation{California Polytechnic State University, San Luis Obispo, CA, 93407, USA.}
\author{H. Schoorlemmer}
\affiliation{Max-Planck-Institut f\"ur Kernphysik, 69117, Heidelberg, Germany.}

\author{W. R. Carvalho Jr.}
\affiliation{Departamento de F\'\i sica,
Universidade de S\~ao Paulo, 
S\~ao Paulo, Brazil.}
\affiliation{Departamento de F\'\i sica de Part\'\i culas \& Instituto Galego de F\'\i sica de Altas Enerx\'\i as, Universidade de Santiago de Compostela, 15782 Santiago de Compostela, Spain.}

\author{J. Alvarez-Mu\~niz} 
\author{E. Zas}
\affiliation{Departamento de F\'\i sica de Part\'\i culas \& Instituto Galego de F\'\i sica de Altas Enerx\'\i as, Universidade de Santiago de Compostela, 15782 Santiago de Compostela, Spain.}

\author{P.~Allison}\atOSU
\author{O.~Banerjee}\atOSU
\author{L.~Batten}\atUCL 
\author{J.~J.~Beatty}\atOSU 
\author{K.~Bechtol}\atUC\atWi
\author{K.~Belov}\atJPL 
\author{D.~Z.~Besson}\atKU\atMoscow
\author{W.~R.~Binns}\atWU 
\author{V.~Bugaev}\atWU 
\author{P.~Cao}\atUD 
\author{C.~C.~Chen}\atNTU 
\author{C.~H.~Chen}\atNTU
\author{P.~Chen}\atNTU 
\author{J.~M.~Clem}\atUD 
\author{A.~Connolly}\atOSU 
\author{L.~Cremonesi}\atUCL 
\author{B.~Dailey}\atOSU 
\author{C.~Deaconu}\atUC 
\author{P.~F.~Dowkontt}\atWU 
\author{B.~D.~Fox}\atUH 
\author{J.~W.~H.~Gordon}\atOSU
\author{P.~W.~Gorham}\atUH 
\author{C.~Hast}\atSLAC
\author{B.~Hill}\atUH 
\author{S.~Y.~Hsu}\atNTU
\author{J.~J.~Huang}\atNTU
\author{K.~Hughes}\atUC\atOSU
\author{R.~Hupe}\atOSU 
\author{M.~H.~Israel}\atWU 
\author{K.~M.~Liewer}\atJPL
\author{T.~C.~Liu}\atNTU 
\author{A.~B.~Ludwig}\atUC 
\author{L.~Macchiarulo}\atUH 
\author{S.~Matsuno}\atUH 
\author{C.~Miki}\atUH 
\author{K.~Mulrey}\atUD\atBRUS
\author{J.~Nam}\atNTU
\author{C.~Naudet}\atJPL
\author{R.~J.~Nichol}\atUCL
\author{A.~Novikov}\atKU\atMoscow
\author{E.~Oberla}\atUC 
\author{S.~Prohira}\atKU
\author{B.~F.~Rauch}\atWU
\author{J.~M.~Roberts}\atUH\atUCSD
\author{B.~Rotter}\atUH
\author{J.~W.~Russell}\atUH 
\author{D.~Saltzberg}\atUCLA
\author{D.~Seckel}\atUD
\author{J.~Shiao}\atNTU
\author{S.~Stafford}\atOSU
\author{J.~Stockham}\atKU
\author{M.~Stockham}\atKU
\author{B.~Strutt}\atUCLA 
\author{M.~S.~Sutherland}\atOSU
\author{G.~S.~Varner}\atUH
\author{A.~G.~Vieregg}\atUC
\author{S.~H.~Wang}\atNTU


\begin{abstract}
Recently, the ANITA collaboration reported on two upward-going extensive air shower events consistent with a primary particle that emerges from the surface of the Antarctic ice sheet. These events may be of $\nu_\tau$ origin, in which the neutrino interacts within the Earth to produce a $\tau$ lepton that emerges from the Earth, decays in the atmosphere, and initiates an extensive air shower. In this paper we estimate an upper bound on the ANITA acceptance to a diffuse $\nu_\tau$ flux detected via $\tau$-lepton-induced air showers within the bounds of Standard Model uncertainties. By comparing this estimate with the acceptance of Pierre Auger Observatory and IceCube and assuming Standard Model interactions, we conclude that a $\nu_\tau$ origin of these events would imply a neutrino flux at least two orders of magnitude above current bounds.
\end{abstract}
\maketitle

\break
\section{Introduction}
\begin{table*}[!t]
  \begin{center}
    \begin{tabular}{l|c|c} 
       & \textbf{ANITA-I: Event 3,985,267} & \textbf{ANITA-III: Event 15,717,147}\\
      \hline
      Payload Elevation Angle & -27.4$^{\circ}\pm 0.3^{\circ}$ & -35.0$^{\circ}\pm 0.3^{\circ}$\\
      Payload Azimuth Angle & 159.6$^{\circ}\pm 0.7^{\circ}$ & 61.4$^{\circ}\pm 0.7^{\circ}$\\
      Payload Altitude & 35.029~km & 35.861~km\\
      Ice Thickness & 3.53~km & 3.22~km\\ 
      Magnetic Field Strength at 0-km & 49.9892~$\mu$T & 60.0783~$\mu$T\\
      Magnetic Field I & -68.24265$^{\circ}$ & -77.4927$^{\circ}$\\
      Magnetic Field D & -38.5059$^{\circ}$ & -155.6842$^{\circ}$\\
      Peak Hpol Electric Field Strength & 0.77 mV/m & 1.1 mV/m \\ 
      Air shower energy & $0.6\pm 0.4$ EeV & $0.6^{+0.3}_{-0.2}$ EeV\\
      \hline
    \end{tabular}
    \caption{\raggedright ANITA-I and ANITA-III $\tau$ candidate events as reported in \cite{ANITA_up} and \cite{ANITA3_tau} respectively. Payload elevation angle refers to the event elevation angle with respect to the payload?s horizontal and payload azimuth angle refers to the event azimuth angle with respect to true north.}
    \label{tab:anita-events-table}	
  \end{center}
\end{table*}
 
The ANITA collaboration has reported the detection of two upward-pointing cosmic-ray-like events propagating directly from the Antarctic ice sheet among a population of $\gtrsim30$ cosmic ray events~\cite{ANITA_up, ANITA3_tau, ANITA_CR}. 
Among the cosmic-ray-like radio signals that reach ANITA from below the horizon, most display a phase reversal indicative of reflections off the ice surface of signals produced by downward-moving Extensive Air Showers (EAS).  However, as described in~\cite{ANITA_up} and~\cite{ANITA3_tau}, ANITA has observed two anomalous events in which radio signals coming from the direction of the ice do not display this phase reversal and thus appear to have been produced by upward-moving EAS. As discussed in~\cite{ANITA_up}~and~\cite{ANITA3_tau}, one plausible mechanism that could produce them is the escape of $\tau$ leptons from $\nu_\tau$ interactions in the Earth and their subsequent decay in the atmosphere to produce an EAS. However, it was noted that the long chord lengths through the Earth pose a severe challenge to this interpretation due to the large probability of absorption~\cite{ANITA_up}. In this work, we explore the hypothesis of $\nu_\tau$ origin within the Standard Model in more detail with an acceptance estimate based on Monte Carlo simulations.

The focus of this work is on estimating the acceptance to a diffuse \nutau flux for comparison with the Auger and IceCube upper limits. We use dedicated particle propagation and EAS radio emission simulations. A follow-up paper will focus on sensitivity to point source fluxes and transients.


Simulations of the radio emission of cosmic-ray EAS using ZHAireS~\cite{ZHAireS} have been applied to interpret the spectral characteristics of the signal~\cite{ZHAireS_UHF}, to predict the effect on signal polarization due to shower charge excess~\cite{ZHAireS_superposition}, and to account for reflections of the radio signals on the ice cap~\cite{ZHAireS_reflected}. The radio emission model in ZHAireS has also been validated in a laboratory experiment that included the effects of a dielectric medium and the influence of a magnetic field~\cite{Belov_2016}. Energy reconstruction of the ANITA-I cosmic-ray-induced events detected after reflection on the ice with ZHAireS has led to the first measurement of the cosmic-ray spectrum with the radio technique~\cite{ANITA_spectrum}, giving compatible results with measurements of the spectrum with more established techniques~\cite{Auger_spectrum, TA_spectrum}. These results give convincing evidence that the simulations of these pulses are accurate.

In this work, we have extended the functionality of ZHAireS to produce EAS radio emission from upward-going $\tau$-lepton decays observed at high altitudes. The simulation allows for the injection of the $\tau$ decay products at any altitude thus enabling the characterization of radio impulsive signals due to $\tau$ decays propagating upwards in the atmosphere. The estimates of the air shower energies presented in~\cite{ANITA_up, ANITA3_tau} used simulations of downward-going cosmic-ray propagation geometries for their interpretation. In this paper, we include the effect of upward-pointing EAS produced at high altitudes. 

We have developed a Monte Carlo simulation to estimate the acceptance of the ANITA instrument to $\tau$-lepton air showers of diffuse $\nu_\tau$ flux origin with the purpose of comparing the sensitivity to the Auger~\cite{Auger_2015, Auger_2017} and IceCube~\cite{IceCube_2016} results, as well as to test whether event emergence angles from the simulations are consistent with the data. The process of producing a $\tau$-lepton decay in the atmosphere from $\nu_\tau$ propagating in Earth is involved and we use publicly available simulations~\cite{Alvarez-Muniz_2018} as part of the acceptance Monte Carlo. On traversal, the $\nu_\tau$ suffers attenuation and regeneration through both neutral and charged current interactions, which, in effect, reduce the neutrino energy.  If a $\nu_\tau$ interaction takes place close to the Earth's surface, it can produce a $\tau$ lepton that travels through the Earth until it exits, with some probability, to the atmosphere. The $\tau$ lepton then decays in flight producing an upward-pointing EAS, which induces a coherent electromagnetic pulse that triggers the ANITA detector floating at high altitude. 

This article is organized as follows. In Section 2 we briefly review the characteristics of the ANITA $\tau$-lepton EAS candidate events and provide results from ZHAireS simulations with the observed geometries for comparison. In Section 3 we provide the details of the acceptance Monte Carlo including an overview of the particle propagation processes involved, the ZHAireS-based radio emission model, and the detector model. Results of the Monte Carlo simulations are presented in Section 4, including the effects of ice shell thickness, Standard Model neutrino-nucleon interaction cross section uncertainties, and two different models of the photonuclear contribution to the $\tau$-lepton energy loss. With this framework, we estimate an upper bound on the ANITA exposure to compare with the $\nu_\tau$ flux limits from Auger and IceCube. In addition, we compare the estimated differential acceptance as a function of emergence angle to the data to test for consistency. In Section 5 we provide discussion and conclusions based on these results.

\section{Radio Emission Modeling of the ANITA $\tau$-lepton air shower candidate events}

\begin{figure}[b!]
  \centering
   \includegraphics[width=0.84\linewidth]{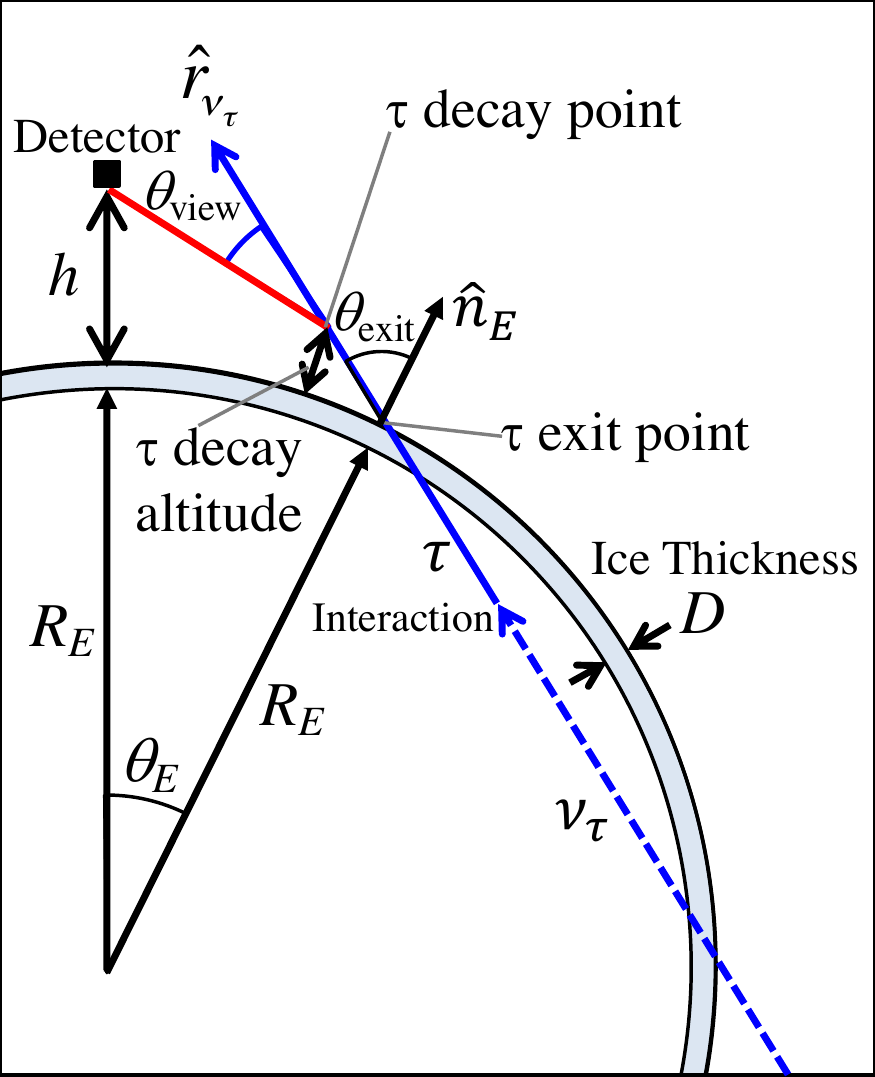} 
   \caption{\raggedright
   Detection geometry. The Earth is modeled as a sphere with Earth's polar radius $R_E$ and a layer of ice of thickness $D$ above that. The detector is at a height $h$ above ice level (i.e. above $R_E+D$). The blue dashed line represents the incoming neutrino with direction of propagation $\hat{r}_{\nu_\tau}$. If the neutrino interacts with the Earth via a charged current interaction, a $\tau$ lepton is produced that continues to propagate with direction $\hat{r}_{\nu_\tau}$. This particle can potentially exit the surface of the Earth at the $\tau$ exit point at Earth angle $\theta_E$. The vector $\hat{r}_{\nu_\tau}$ is not necessarily in the plane of the page. The exit angle $\theta_\mathrm{exit}$ is the angle between the vector normal to the surface of the Earth at the exit point $\hat{n}_{E}$ and $\hat{r}_{\nu_\tau}$. If a tau lepton exits the surface of the Earth, it will propagate in the Earth's atmosphere until it decays at a $\tau$ decay altitude above the ice surface (i.e. above $R_E+D$). If the decay mode includes hadrons, it will produce an extensive air shower (EAS). This EAS will produce a radio impulse. Because in some cases the shower maximum can be near or past the location of the detector, the view angle $\theta_\mathrm{view}$ of the radio emission is taken with respect to the $\tau$-lepton decay point. }
   \label{fig:geom}
\end{figure}

Event 3,985,267 from ANITA-I~\cite{ANITA_up} and event 15,717,147 from ANITA-III~\cite{ANITA3_tau} are isolated events that passed all signal quality and clustering cuts. The electric fields are impulsive and have spectra consistent with the other ANITA cosmic ray events~\cite{ANITA_CR, ANITA_spectrum} and their polarizations are correlated with the geomagnetic field. The distinguishing feature is that the polarity, the sign of the maximum electric field value, of these events is inverted compared to the rest of the cosmic ray events pointing to the continent. This is a feature that is consistent with the radio emission of an extensive air shower that is not reflected. Interpreting these events as extensive air showers requires that the parent particles producing them emerge upward from the ice, particularly because the measured emergence angles (the complement of the exit angle $\theta_\mathrm{exit}$ shown in Figure~\ref{fig:geom}) of the events are 25.4$^\circ$ (ANITA-I) and 35.5$^\circ$ (ANITA-III) with $\sim1^\circ$ uncertainty. We summarize the event parameters in Table~\ref{tab:anita-events-table}.

These upgoing showers could be due to a tau neutrino incident on the Earth. The $\nu_\tau$ would have to propagate through most of the matter depth, either directly or with regeneration, before producing a $\tau$ lepton via a charged-current interaction near the surface, with the $\tau$ lepton subsequently decaying in the atmosphere and at least one of its decay products initiating an extensive air shower. When assuming the ANITA events are due to $\tau$-lepton decay, we must consider the decay location in the atmosphere. The $\tau$ lepton decay range is $L\sim(\mathcal{E}_\tau/\rm{EeV})\times49~\rm{km}$ with ${\mathcal{E}_\tau}$ the energy of the $\tau$, meaning that the event could have decayed tens of km further along its trajectory in the atmosphere after exiting the ice. 

The geometry for detecting tau lepton air showers from neutrinos piercing the Earth is shown in Figure~\ref{fig:geom}. 
If a tau neutrino enters the surface of the Earth, it may produce a tau lepton that exits the surface of the Earth at the other end. 
A tau lepton propagating into the atmosphere will eventually decay with a rest-frame lifetime $2.9\times10^{-13}$~seconds. 
The $\tau$ lepton will decay into a hadronic mode with a probability of 64.8\% \cite{tauola}, thus producing an extensive air shower. The radio emission of such a shower could be observed by a receiver at altitude $h$.

\label{sec:review}

\begin{figure*}[!ht]
\includegraphics[width=0.49\textwidth]{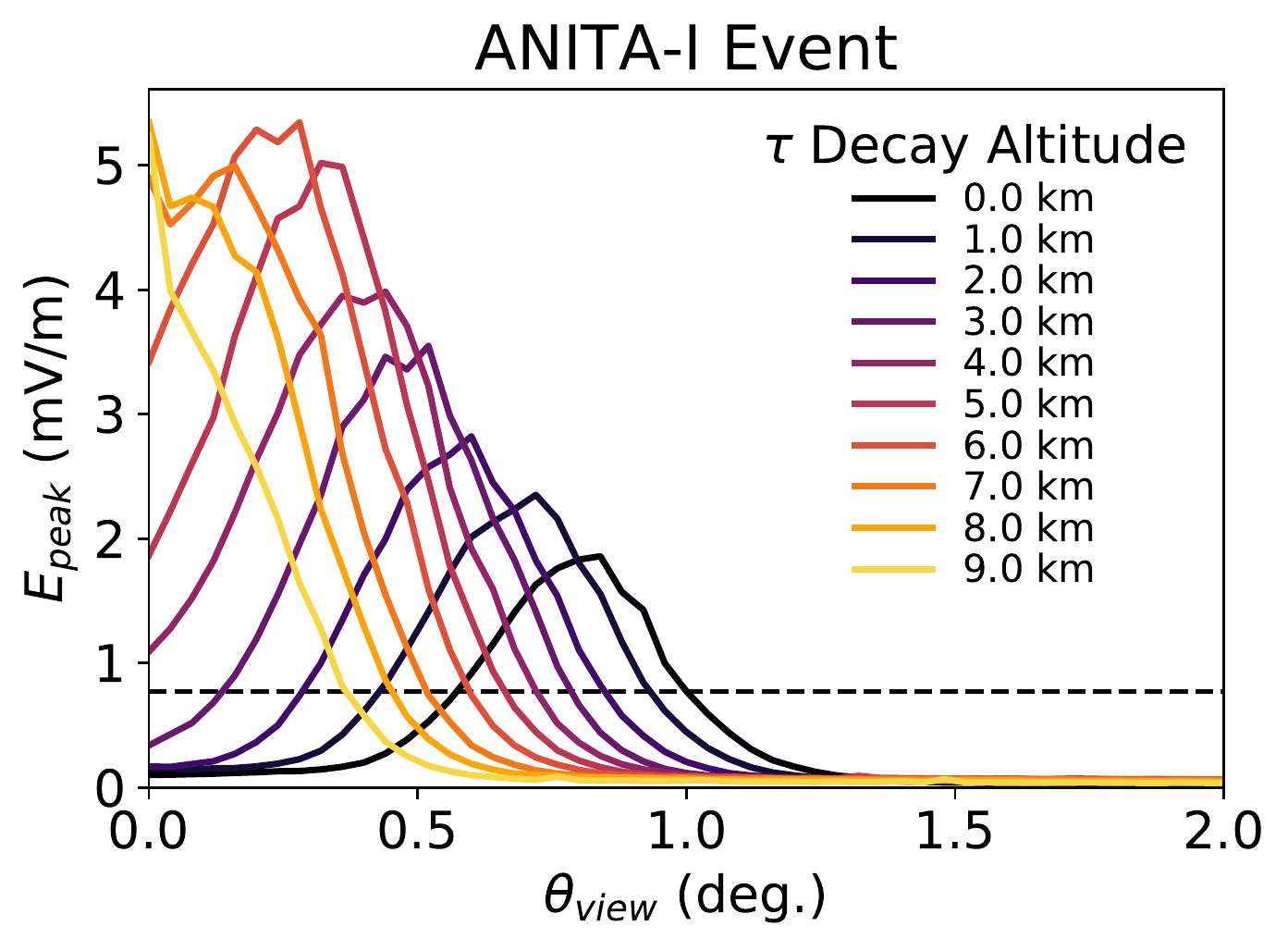} 
\includegraphics[width=0.49\textwidth]{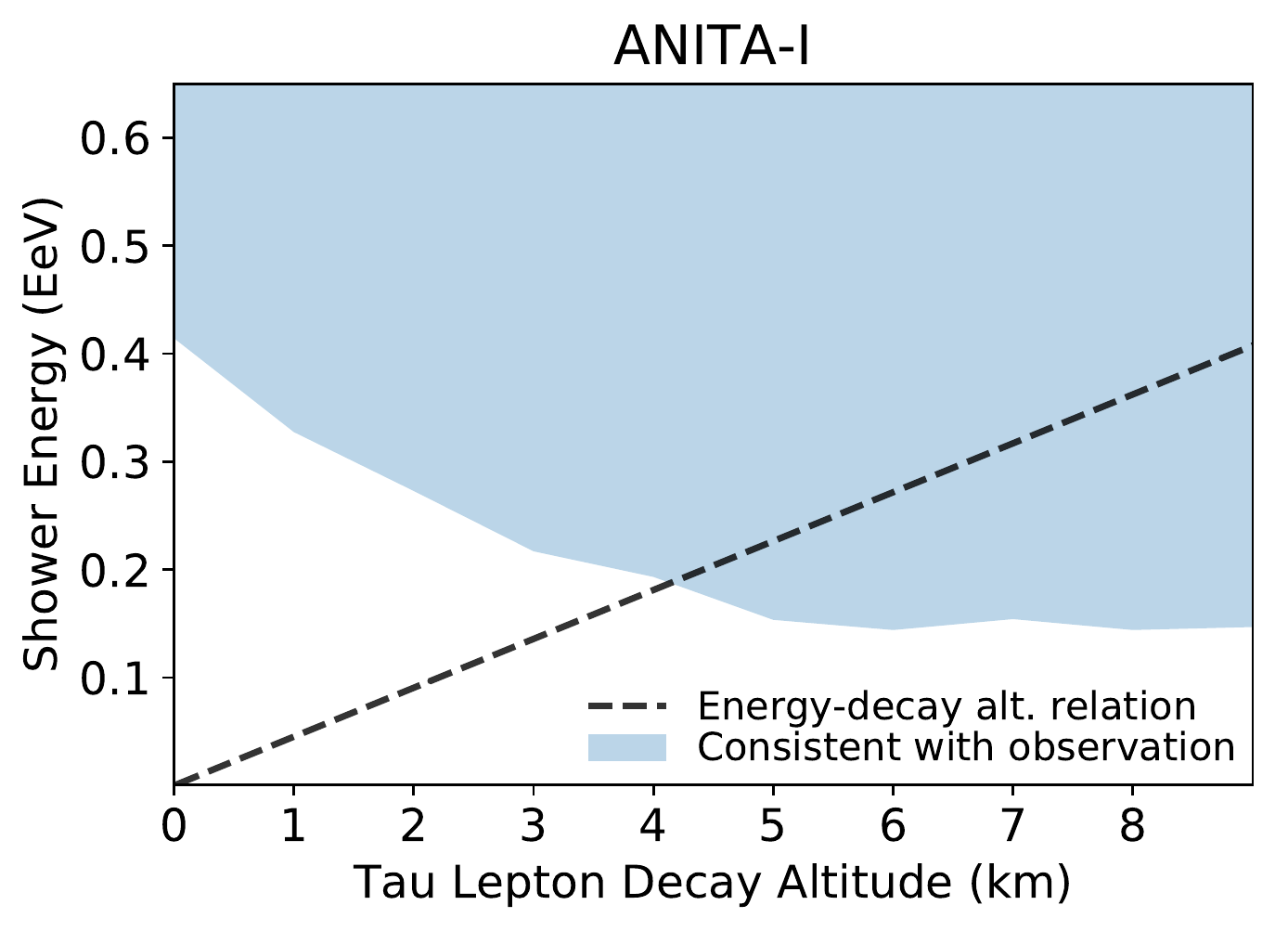}
\includegraphics[width=0.49\textwidth]{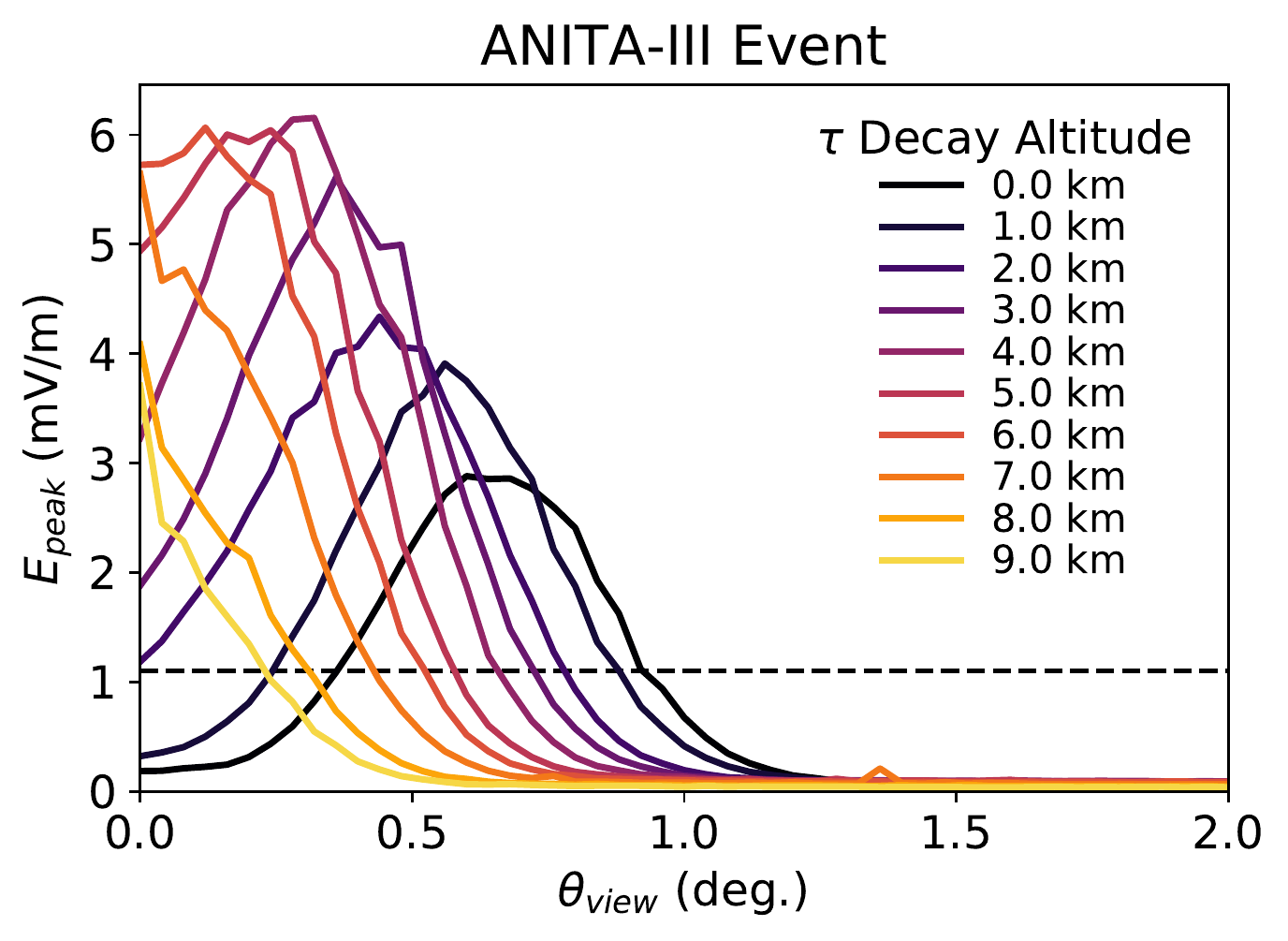} 
\includegraphics[width=0.49\textwidth]{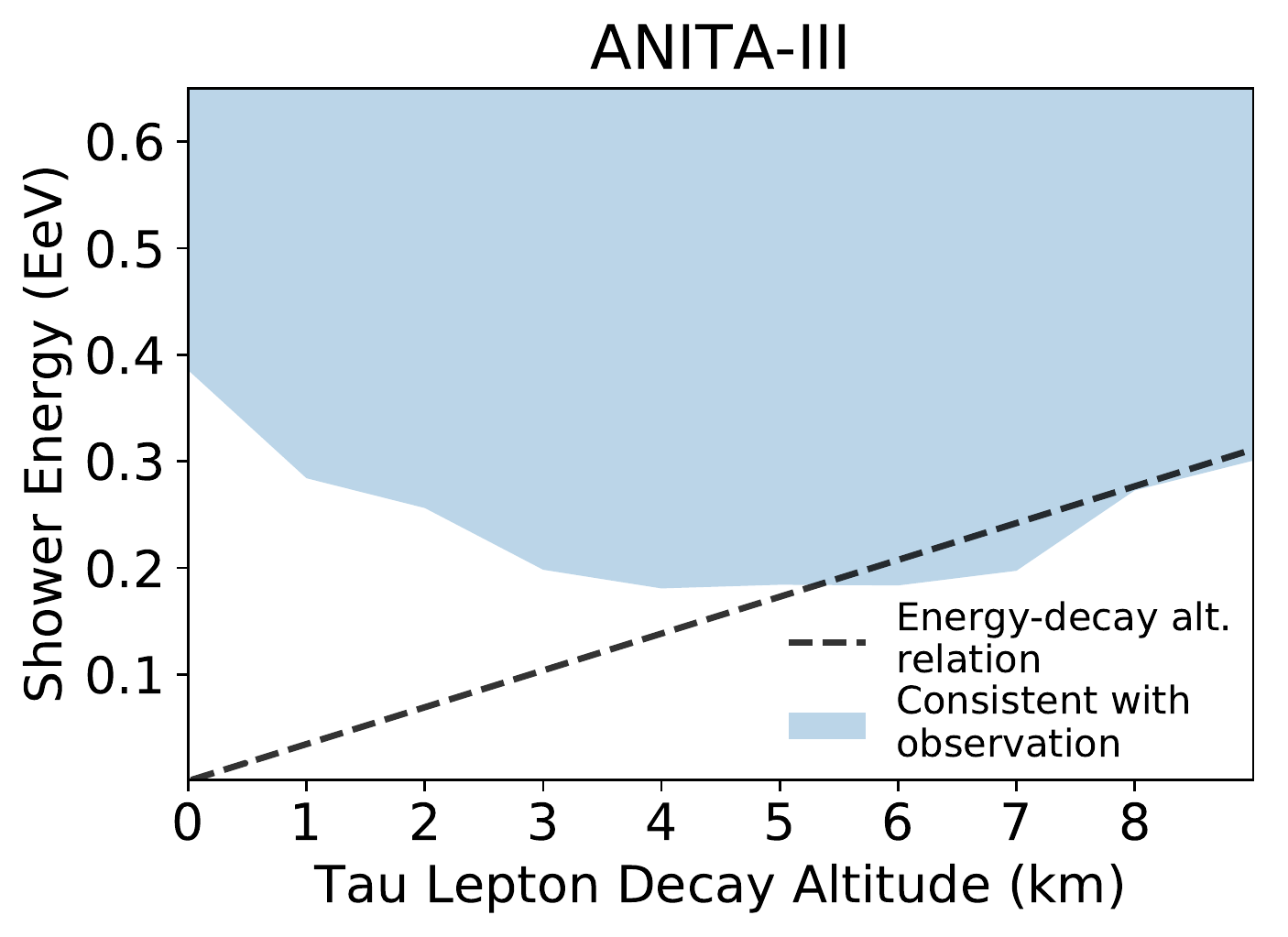}
\caption{\raggedright 
Left: The simulated peak electric field strength from ZHAireS at the ANITA payload 
as a function of view angle $\theta_{\rm view}$ with respect to the shower axis at the $\tau$-lepton decay point (see Figure~\ref{fig:geom}),
for varying decay altitudes for both the ANITA-I (top) and ANITA-III events (bottom). We assume the magnetic field strength, inclination angle and event parameters  
from Table~\ref{tab:anita-events-table}, ice thicknesses of 3.0~km, and a bandwidth of 180-1200~MHz. The simulated shower energy is $10^{18}$~eV.
The horizontal dashed lines mark the measured electric field strength of the events at the ANITA payload. 
Right: 
The minimum shower energy, at a given $\tau$-lepton decay altitude, that results in a peak electric field consistent with values observed by ANITA. The minimum shower energy is obtained by scaling the peaks of the radio emission profiles shown on the left with energy.
Shower energies above this minimum are shown in the shaded blue regions.
With the decay channel assumed (described in Section \ref{sec:radio_emission_param}), the shower energy is roughly equal to the energy of the tau lepton. 
The dashed line is the energy-decay altitude relation given by converting the decay range $L\sim(\mathcal{E}_\tau/\rm{EeV})\times49~\rm{km}$ to altitude. 
}


\label{fig:energydecayheight}
\end{figure*}

In Figure~\ref{fig:energydecayheight} we show a set of radio emission profiles from air showers initiated at different decay altitudes. These profiles were simulated with ZHAireS~\cite{ZHAireS} using the geomagnetic fields in Table~\ref{tab:anita-events-table} adapted to the upward-going air shower geometries and bandwidths corresponding to the ANITA events. 
The peak electric field for each decay altitude defines the minimum energy of the observed showers, shown in the right panels of Figure \ref{fig:energydecayheight}. Changes in the radio emission profile at higher altitudes result in variations in the shower energy estimate. The electric field at the peak increases with $\tau$ decay altitude up until $\sim 5$~km, because the shower maximum moves closer to the detector. Above 5~km, the peak decreases with altitude because the air shower is not fully developed.  
We estimate that the tau shower energy at 0~km decay altitude above ice level is 0.67 EeV for the ANITA-I event and 0.56 EeV for the ANITA-III event, consistent with prior estimates scaled from downward-going cosmic-ray air showers~\cite{ANITA_spectrum}. However, as shown in Figure \ref{fig:energydecayheight}, the lack of knowledge of the tau decay altitude leads to a factor $\sim 2$ uncertainty on the tau shower energy. 
The shower energy uncertainty reported by ANITA-I of $0.6\pm0.4$~EeV is larger than the uncertainty due to decay altitude while the ANITA-III reported uncertainty $0.56^{+0.3}_{-0.2}$~EeV has a smaller lower bound than expected from decay altitude alone.
The uncertainty in the view angle also contributes to the uncertainty in the shower energy, although this in principle can be further constrained using the spectral slope of the radio emission~\cite{ANITA_spectrum}.

The minimum shower energy for these events is obtained for tau decay altitudes above 4~km. This altitude is consistent with that expected for typical tau decays of roughly the same energies for both events as indicated by the dashed line in Figure \ref{fig:energydecayheight}. The consistency among the observed electric fields, shower energies, and expected tau decay altitudes is not discrepant with the upward-going $\tau$ lepton hypothesis.

\section{Acceptance Estimate}


In this section we present the model used for our estimates of the ANITA acceptance to $\tau$ lepton air showers of $\nu_\tau$ origin. 
The acceptance estimate relies on $\tau$ neutrino propagation, radio emission model, and the detector model. The goal of this work is to provide an upper bound of the acceptance and compare it to other experiments. Several approximations are taken along the way to simplify the estimate. We make optimistic approximations while keeping them at the relevant scale. Note that in this section we are no longer characterizing the ANITA-I and ANITA-III events but rather providing an estimate of the ANITA acceptance.

The acceptance to a diffuse flux $\langle A\Omega\rangle$ is given, differentially, by $d\langle A\Omega\rangle=dA \ d\Omega_\nu \ \hat{n}\cdot \hat{r}_\nu \ P_\mathrm{obs}$. The differential area $dA$ with normal vector $\hat{n}$ is a reference region for the passage of a flux of particles. The direction of the particle axis of propagation is given by $\hat{r}_\nu$ with differential solid angle $d\Omega_\nu$. The dot product accounts for the projected area in the direction of the particle. $P_\mathrm{obs}$ is the probability that a particle axis of propagation passing through the reference area element $dA$ with direction $\hat{r}_\nu$ is observed. This includes all attenuation factors, production of the observable electromagnetic waves, and detection as discussed below. 

For the ANITA observation geometry, the natural choice of reference area is the surface of the Earth including the ice layer. To simplify the problem, we take the area of integration to be the spherical cap visible from the detector at altitude $h$ above ice level (not sea level), the assumption being that a particle entering the surface of the Earth must exit the surface visible to the detector to produce an air shower visible at high altitude. This is neglecting a small region beyond the horizon where a $\tau$-lepton could exit and decay after many km of propagation producing a potentially detectable signal. However, since we are interested in relatively high emergence angles and lower energies, where the $\tau$ decay range in the atmosphere is $\lesssim50~\rm{km}$ (given the ANITA events of interest) we do not include this possibility, although it could be added to future estimates. 
The probability of observation $P_\mathrm{obs}$ includes multiple components.  The first is the probability that a $\tau$ lepton exits the ice into the atmosphere. This must also account for the distribution of energies $\mathcal{E}_\tau$ of the lepton exiting the ice given the parent neutrino energy $\mathcal{E}_\nu$, which we denote as $p_\mathrm{exit}(\mathcal{E}_\tau|\mathcal{E}_\nu, \theta_\mathrm{exit})$. 

The $\tau$ lepton subsequently decays in the atmosphere after an exponentially distributed distance $s_\mathrm{decay}$ leading to $p_\mathrm{decay}(s_\mathrm{decay}|\mathcal{E}_\tau)=\exp\left(-s_\mathrm{decay}/L(\mathcal{E}_\tau)\right)$ with $L(\mathcal{E}_\tau)/\mathcal{E}_\tau=49~\rm{km}/\rm{EeV}$. There is the possibility that the decay takes place past the detector, which increases with increasing energy. These events do not contribute to the total acceptance. 

Upon decay, the daughter particles will interact with the atmosphere to produce an air shower. The most common $\tau$-lepton decay mode results in $\pi^-\pi^0\nu_\tau$ with most of the energy ($\sim$98\%) going into the pions, which produce an extensive air shower. In this work, this is the injected set of particles used for shower simulations. In general, the energy going into an extensive air shower $\mathcal{E}_\mathrm{EAS}$ given $\mathcal{E}_\tau$ has a probability density function $p_\mathrm{EAS}(\mathcal{E}_\mathrm{EAS}|\mathcal{E}_{\tau})$.
For our upper bound estimate we take the optimistic assumption that $\mathcal{E}_\mathrm{EAS}=\mathcal{E}_\tau$, which is close to within a few percent for the most common $\tau$-lepton decay mode. This would have to be treated in more detail for a higher fidelity estimate, including the $\tau$-lepton decay modes that produce no hadrons. 

The shower then produces radio impulsive emission with peak electric field $E_{\rm peak}$ at the location of the detector with a probability density function 
$p_\mathrm{pk}(E_{\rm peak}|\mathcal{E}_\mathrm{EAS}, s_\mathrm{decay}, \hat{\mathbf{r}}_{\nu_\tau}, \mathbf{x}_\mathrm{det}, \mathbf{x}_\mathrm{exit})$.
The radio impulse spectrum and strength at the payload depend on distance and view angle $\theta_{\rm view}$, which is the angle between the shower axis and the line joining the detector position and the $\tau$ decay point (see Figure \ref{fig:geom}), as well as the atmospheric density profile in which the air shower develops.
This is accounted for by keeping track of the decay position.
The distance and $\theta_{\rm view}$ are determined by the exit point $\mathbf{x}_\mathrm{exit}$, position of the detector $\mathbf{x}_\mathrm{det}$, direction of propagation $\hat{\mathbf{r}}_{\nu_\tau}$, and decay distance $s_\mathrm{decay}$.
For this acceptance estimate, we produce radio emission profiles for a range of decay altitudes and $\tau$ lepton propagation directions (emergence angles). These are parameterized (Section III.B) for use in a Monte Carlo evaluation of the acceptance.
Finally,
the probability that the detector triggers $p_\mathrm{trig}(E_\mathrm{peak}$) depends on the peak electric field and beam pattern of the antennas.
The acceptance of tau neutrinos, including all the steps described above, is given by the nested integral
%
\begin{equation}
\begin{split}
\langle A\Omega\rangle_{\nu_\tau}(\mathcal{E}_{\nu_\tau})
= 
& R_E^2 \iint d\Omega_E \iint d\Omega_{\nu_\tau} \ \hat{r}_{\nu_\tau}\cdot \hat{n}_E 
\\
& 
\int d\mathcal{E}_{\tau } \ p_{\mbox{\scriptsize exit}}(\mathcal{E}_{\tau}|\mathcal{E}_{\nu_\tau}, \theta_{\mbox{\scriptsize exit}}) 
\\
& 
\int ds_\mathrm{decay} \ p_\mathrm{decay}(s_\mathrm{decay}|\mathcal{E}_{\tau}) 
\\
&
\int d\mathcal{E}_\mathrm{EAS} \ p_\mathrm{EAS}(\mathcal{E}_\mathrm{EAS}|\mathcal{E}_{\tau}) 
\\
& 
\int dE_\mathrm{peak} \ p_\mathrm{pk}(E_\mathrm{peak}|\mathcal{E}_\mathrm{EAS}, s_\mathrm{decay}, \hat{\mathbf{r}}_{\nu_\tau}, \mathbf{x}_\mathrm{det}, \mathbf{x}_\mathrm{exit}) 
\\
&
\ \ \ \ p_\mathrm{trig}(E_\mathrm{peak}) \\
\end{split}
\label{eq:acceptance}
\end{equation}

The surface integral is performed over the surface of a spherical Earth model with polar radius, $R_E=6,356.7523$~km and differential solid angle, $d\Omega_E$, with polar coordinates $\theta_E$, $\phi_E$ (see Figure~\ref{fig:geom}). The normal vector to the Earth's surface at the tau lepton exit point is $\hat{n}_{E}$. The solid angle integration about the neutrino directions is $d\Omega_{\nu_\tau}$, in polar coordinates defined locally at the exit point, with $\theta_\nu$ referenced to $\hat{n}_{E}$ and $\phi_\nu$ referenced to the direction to the payload. 

In the following subsections, we provide details of the $\tau$ neutrino and lepton propagation, radio emission model, and detector model, including discussion of the approximations used for the upper bound estimate of the acceptance. 



\subsection{$\tau$ neutrino and lepton propagation}
\label{sec:tau_exit_prob}

For the evaluation of $p_{\mbox{\scriptsize exit}}(\mathcal{E}_{\tau}|\mathcal{E}_{\nu_\tau}, \theta_{\mbox{\scriptsize exit}})$ we use the publicly available propagation code~\cite{Alvarez-Muniz_2018}. This code allows the user to specify different ice thicknesses, Standard Model neutrino-nucleon cross sections, and $\tau$-lepton energy loss models. We include calculations using different possibilities for these effects in the results of this paper. 

The $\tau$ exit probabilities, marginalized over the exiting $\tau$ lepton energy, are given by: 
\begin{equation}
P_{\mbox{\scriptsize exit}}(\mathcal{E}_{\nu_\tau}, \theta_{\mbox{\scriptsize exit}})=\int d\mathcal{E}_\tau ~p_{\mbox{\scriptsize exit}}(\mathcal{E}_{\tau}|\mathcal{E}_{\nu_\tau}, \theta_{\mbox{\scriptsize exit}})
\label{eq:exitprob}
\end{equation}
These probabilities have been characterized in detail in~\cite{Alvarez-Muniz_2018} where the $\mathcal{E}_\tau$ distributions are provided as well. 

The main results presented in~\cite{Alvarez-Muniz_2018} relevant to this study are listed as follows. The effect of $\nu_\tau$ regeneration, where a neutrino interacts in the Earth via a charged-current interaction producing a $\tau$ lepton that subsequently decays into a  lower energy $\nu_\tau$ is important at emergence angles $>3^\circ$. Not including it severely underestimates the sensitivity to $\nu_\tau$ for observatories at high altitudes, such as ANITA. The presence of a layer of ice $>$ 1~km thick results in an increased $P_{\mbox{\scriptsize exit}}$ compared to bare rock only for $\nu_\tau$ energies above $3\times10^{18}$~eV. Below this energy, the presence of an ice or water layer reduces $P_{\mbox{\scriptsize exit}}$ due to the low probability of a neutrino interaction compared to the reduced $\tau$-lepton decay range. 
Finally, it was also found that for emergence angles $\gtrsim 5^{\circ}$, the Earth acts as a filter reducing the high energy $\tau$-lepton flux. 
This is the regime where regeneration dominates the outgoing flux of $\tau$ leptons.

\subsection{Radio emission model}
\label{sec:radio_emission_param}




We model the radio emission from a particle cascade initiated by the decay of an ultra-high-energy $\tau$ lepton using the \ZHAireS code~\cite{ZHAireS}. This code implements the ZHS algorithm \cite{ZHS92,Garcia-Fernandez2013}, which calculates the total radio signal by summing the emission from each single particle track obtained from the AIRES~\cite{ZHAireS} simulation for atmospheric particle cascades. To initialize the particle shower, we feed into AIRES the products of a $\tau$-lepton decay, obtained from TAUOLA \cite{tauola} simulations of tau decays at several energies. The energy of the products of a specific decay can be scaled to obtain a specific $\tau$ energy or shower energy. These decay products are injected into the atmosphere at the desired decay altitude. By propagating these decay products, ZHAireS creates the atmospheric shower and calculates the radio emission.

  In the radio simulations shown in this work we used a single TAUOLA simulated decay at $10^{17}$~eV, with the most common (25\%) $\tau$-lepton decay mode ($\pi^-\pi^0\nu_\tau$). In this simulation, the three decay products take 67\%, 31\%, and 2\% of the original $\tau$-lepton energy, respectively. 

For this study we developed a special version of the ZHAireS code, capable of correctly handling time calculations for up-going showers starting anywhere in the atmosphere. This makes it possible to freely choose the location of the decay as well as the direction of propagation for the $\tau$ decay products. 

\begin{figure}[!htbp]
\centering
\includegraphics[height=0.39\textwidth]{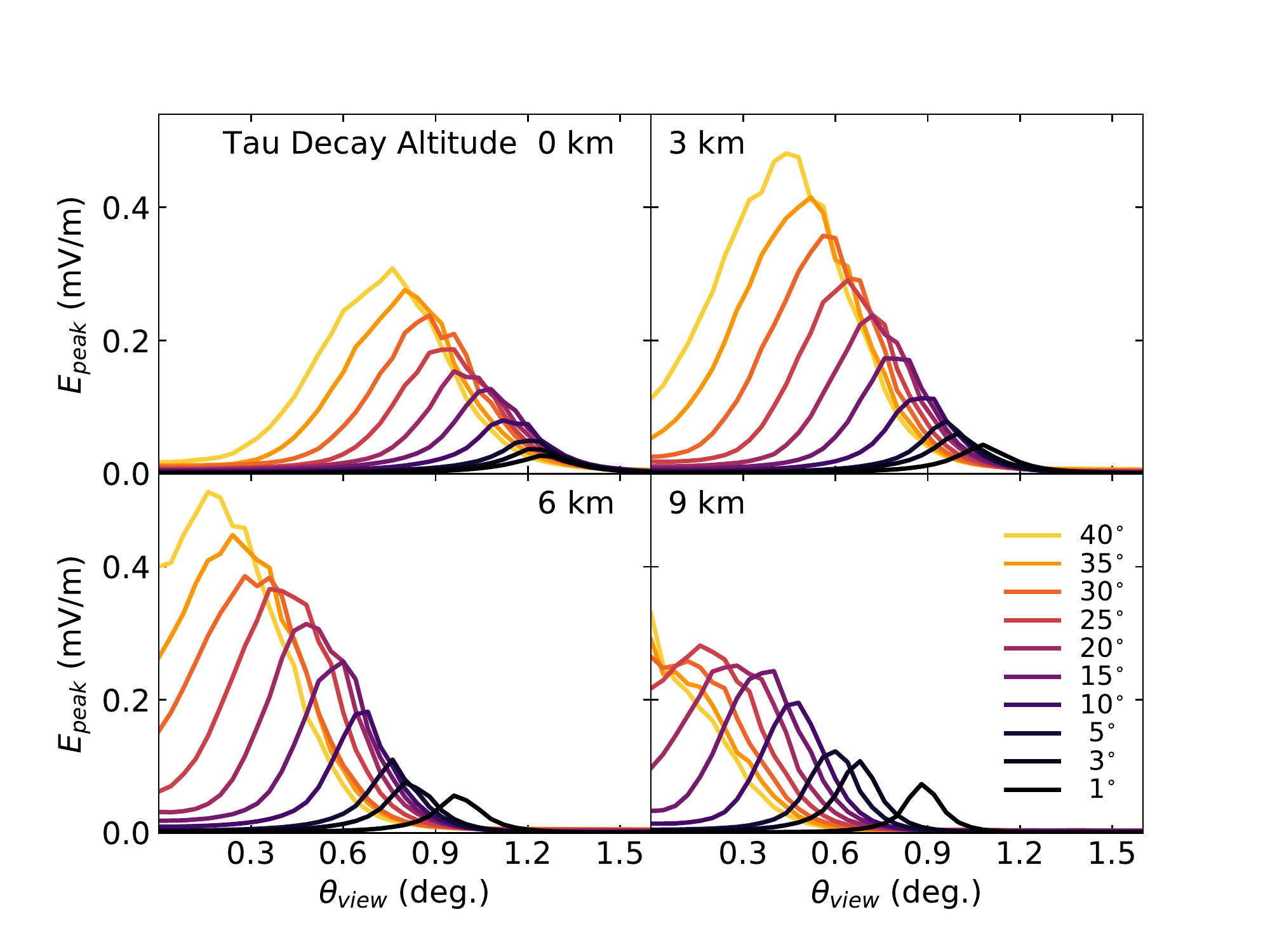} 
\centering
\includegraphics[height=0.375\textwidth]{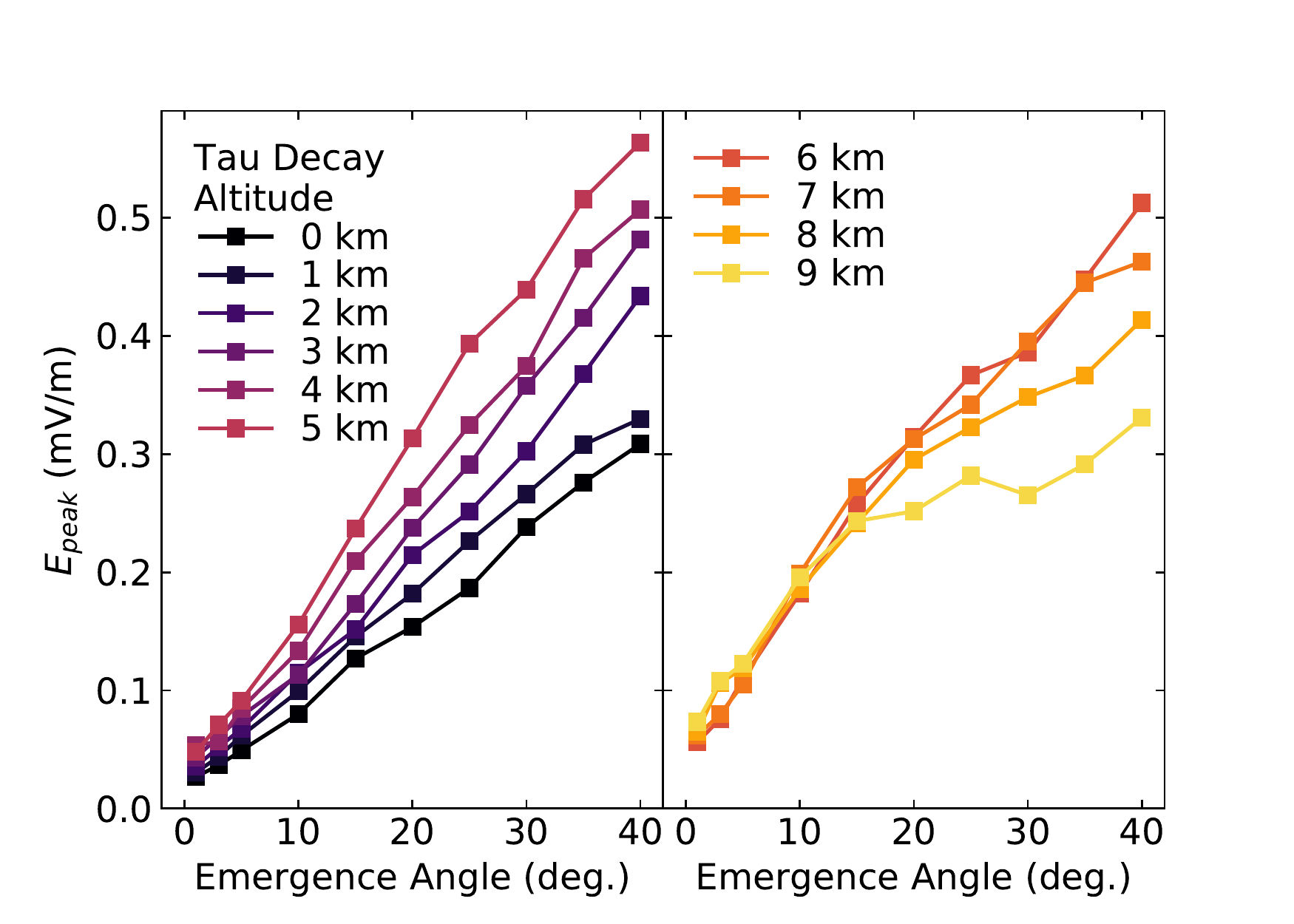}
\caption{\raggedright 
The peak electric field resulting from $10^{17}$\,eV $\tau$-lepton decay as a function of view angle $\theta_\mathrm{view}$ with respect to the shower axis at the $\tau$-lepton decay point (see Figure~\ref{fig:geom} for geometry). Top panels: Electric field profiles are shown for $\tau$ decay altitudes from 0~km (ice level) to 9~km and emergence angles from 1$^{\circ}$  to 40$^{\circ}$. The observation point is at 37~km altitude. Note that the emergence angle is measured relative to ice level while the view angle is measured relative to a shower at the decay point. Bottom panels: the peak value of the profiles as a function of emergence angle. See text for explanation. 
}
\label{fig:tau_altitude_and_emergence}
\end{figure}


For the acceptance estimate portion of this study, we simulated showers with a magnetic field of 60~$\mu$T. In each case, the magnetic field vector is oriented perpendicular to the direction of the shower. The electric field is filtered in the 180-1200~MHz band to match the trigger band of ANITA-III. This produces the largest possible emission for our upper bound estimate. In Figure \ref{fig:tau_altitude_and_emergence}, we show simulated peak electric fields, filtered in the 180-1200~MHz band, as a function of view angle ($\theta_\mathrm{view}$) with respect to the $\tau$ lepton decay point for $\mathcal{E}_\tau=10^{17}$~eV and for various decay altitudes and emergence angles.
%


Different stages of the shower contribute with varying levels of coherence to the total electric field depending on distance to observer, number of particles, and emission angle. As the shower develops the angle of the line of sight to the observation point changes introducing time delays which can result in constructive or destructive interference between different stages in the longitudinal development of the shower. Also, as the detector moves away from the shower axis, the distance to the emission region changes resulting in additional time delays \cite{ZHAireS_UHF}. The net result is a ring-like radio emission pattern as shown in the top panel of Figure~\ref{fig:tau_altitude_and_emergence} with a maximum at a certain viewing angle $\theta^\mathrm{max}_{\rm view}$ relative to the shower axis as seen from the $\tau$-decay point.

\begin{figure}[htbp]
  \centering
\includegraphics[width=0.5\textwidth]{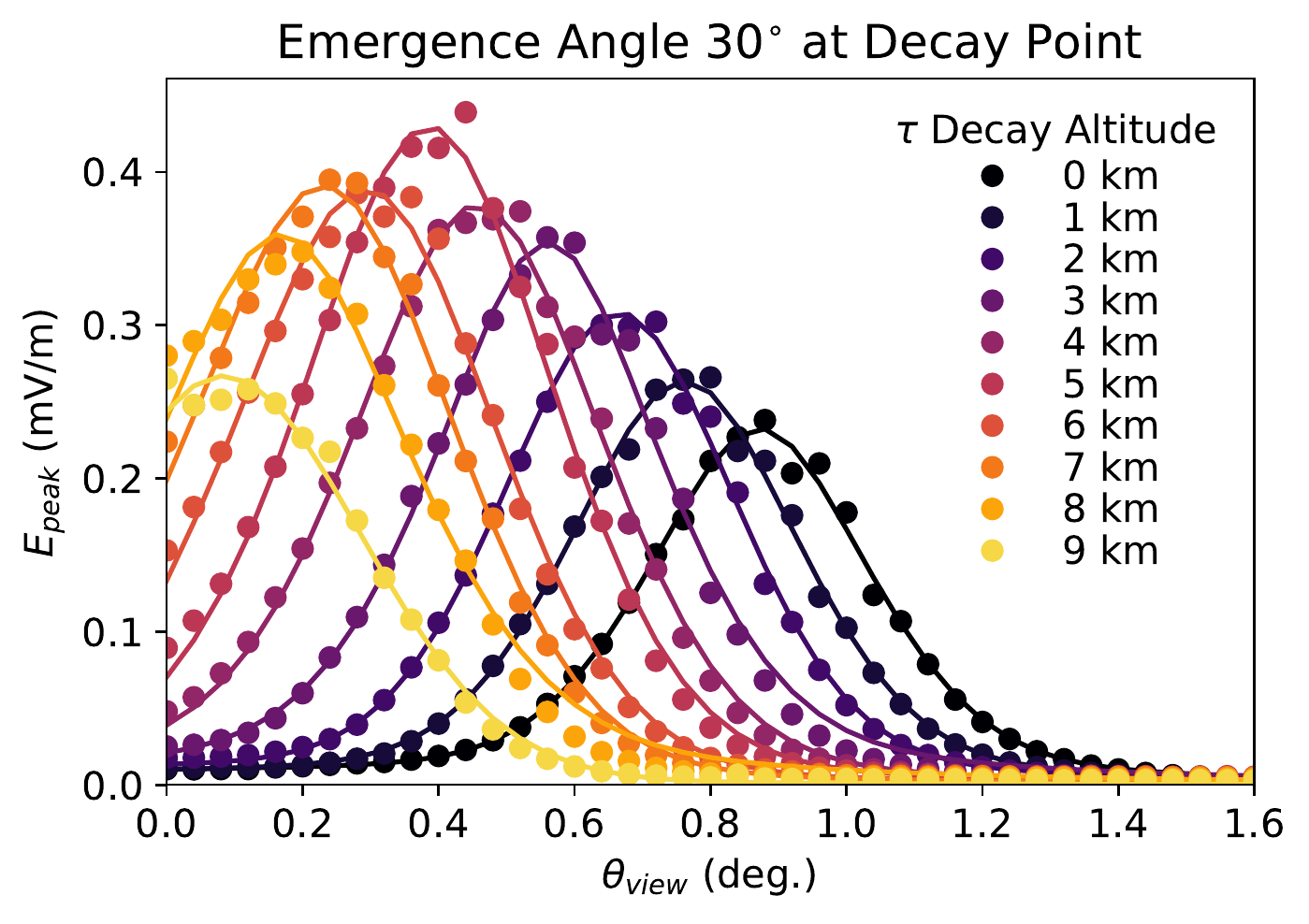} 
\caption{\raggedright 
The peak electric field vs. $\theta_\mathrm{view}$ from ZHAireS simulations of a $10^{17}$\,eV $\tau$ lepton decay at emergence angle 30$^{\circ}$ and for decay altitudes from 0~km to 9~km (dots) compared with the parameterized fits (lines) described in Eqs. (\ref{eq:param1}) and (\ref{eq:param2}).}
\label{fig:zhaires_fits}
\end{figure}

For $\tau$-lepton decays at low altitudes, the induced showers reach $X_{\rm max}$ before ANITA and $\theta^\mathrm{max}_\mathrm{view}$ roughly corresponds to viewing an extended region around $X_\mathrm{max}$ at angles close to the Cherenkov angle where the coherence is maximal \cite{ZHAireS_UHF}. For $\tau$-lepton decays at high altitudes $X_\mathrm{max}$ is reached past ANITA. For instance at a decay altitude $\sim$6~km above the ice,
a 30$^{\circ}$ shower of energy 0.5~EeV reaches on average its maximum size around the detector position. 
In this case there is a competition between an increase in electric field due to 
the reduced distance between the shower and the observer,
and a decrease in signal strength due to the shower evolving in a thinner atmosphere and not fully developing before reaching ANITA with only a small fraction of the early shower development contributing to the coherent pulse. This results in weaker signals despite the shower being closer to the detector. Also the beam narrows because of geometric projection effects due to the Cherenkov emission conical beam pattern produced along shower development starting closer to ANITA. The beam narrows further due to the refractivity scaling (to first order) with the atmospheric density and hence the Cherenkov angle decreasing with altitude. These trends can be clearly observed in Figure~\ref{fig:zhaires_fits}, where we show the radio emission profiles at fixed emergence angle of 30$^\circ$ for various decay altitudes.

Since the full radio simulation of EAS is computationally intensive, we parameterize the behavior at discrete values of shower parameters. For a given shower in the acceptance estimate, we use the parameterization associated with the decay altitude and emergence angle nearest to the shower geometric variables and scale the electric field amplitude linearly with EAS energy.

For each $\tau$-decay altitude and emergence angle, we parameterize the radio emission beam pattern in the 180-1200 MHz band for the Monte Carlo estimate of the acceptance. The functional form of the fits is given by the combination of a Gaussian and Lorentzian centered on the peak $\theta_{\rm view}$ along with a
Gaussian centered at $\theta_\mathrm{view}=0^{\circ}$. The shape is given by
\begin{equation}
\begin{split}
\epsilon(\theta_\mathrm{view}) = & E_0\left[
f\exp\left(-\frac{(\theta_\mathrm{view}
    - \theta_\mathrm{pk})^2}{2\sigma_\mathrm{view}^2}\right) + \right.
    \\
    &
    \left.
  (1-f)\left(1+\left(\frac{\theta_\mathrm{view} -
    \theta_\mathrm{pk}}{\sigma_\mathrm{view}}\right)^2\right)^{-1}\right] +
    \\
    &
E_1\exp\left(-\frac{\theta_\mathrm{view}^2}{2\Sigma_\mathrm{view}^2}\right) 
\end{split}
\label{eq:param1}
\end{equation}
and the electric field is
\begin{equation}
E_\mathrm{peak}(\mathcal{E}_{\tau}, r, \theta_\mathrm{view}) = \left(\frac{\mathcal{E}_{\tau}}{10^{17}\mbox{
    eV}}\right)\left(\frac{r}{r_0}\right)^{-1}\epsilon(\theta_\mathrm{view})
\label{eq:param2}
\end{equation}
where $r_0$ is the distance from ANITA to the $\tau$ lepton decay point. Note that $r_0$ is not a free parameter of the fit but rather it just varies depending on the chosen decay altitude and emergence angle. The parameterization is done for emergence angles of $1^{\circ}$, $3^{\circ}$, and $5^\circ-40^\circ$ in $5^\circ$ degree steps as well as decay altitudes in the range 0-9~km in 1~km steps.
%
As an example, the best fit parameters for an emergence angle of 30$^{\circ}$ and decay altitude of 0~km are $E_0=0.151$~mV/m,
$\theta_\mathrm{pk}=0.873^{\circ}$, $\sigma_\mathrm{view}=0.161^{\circ}$, $f=0.745$,
$E_1=1.549\,\mu$V/m, $\Sigma_\mathrm{view}=0.176^{\circ}$. The parameterization of the peak electric field as a function of view angle for various $\tau$-lepton decay altitudes, along with the simulated points, are shown in Figure~\ref{fig:zhaires_fits}. We have verified that the simulation correctly reproduces the tails of the emission beam pattern to within 4\%.

\subsection{Detection model}
\label{sec:simple_ANITA_model}


The calculation of the probability of detection must account for the position of the tau decay in the atmosphere ($p_\mathrm{decay}$ in Equation~\ref{eq:acceptance}), the production of the extensive air shower ($p_\mathrm{EAS}$), its radio emission 
($p_\mathrm{pk}$), 
and the detector trigger ($p_\mathrm{trig}$). 
The shower initiation point 
$s_\mathrm{decay}$
with respect to the exit point along the neutrino axis of propagation is sampled with an exponential distribution  
$p_\mathrm{decay}(s_\mathrm{decay})=\exp(-s_\mathrm{decay}/L)$
where $L$ is the $\tau$-lepton decay range. 
The probability that the shower is hadronic $P_{\mbox{\scriptsize hadron}}=64.8\%$ is taken into account in $p_\mathrm{EAS}$.
The energy $\mathcal{E}_\mathrm{EAS}$ is 98\% of $\mathcal{E}_\tau$ based on the decay mode assumed (see Section 3.2) and we assume all the energy of the $\tau$ lepton goes into producing an extensive air shower, so that the integral in $\mathcal{E}_\mathrm{EAS}$ can be omitted setting $\mathcal{E}_\mathrm{EAS}=\mathcal{E}_\tau$.

The ANITA-I trigger model is fully described in~\cite{ANITA_instrument_paper}. Each antenna consists of two linearly polarized channels. The signals are combined into two circular polarizations and split into four sub-bands per polarization. For an antenna to trigger, three of eight sub-bands must be above threshold. The exponentially falling spectrum of extensive air shower radio emission at frequencies above 300 MHz means that while lower frequency ANITA sub-bands may exceed the thermal thresholds, the higher frequency sub-band may not. Overall, this results in a higher threshold over the full band. 

The ANITA-III instrument was updated to include a full-band impulsive trigger, additional antennas, and lower noise amplifiers. However, persistent continuous wave radio-frequency interference from satellites in the North were masked out from consideration in the trigger (a feature called phi-masking), resulting in a decreased exposure.  Details of the ANITA-III trigger and performance are available in~\cite{ANITA-3-askaryan, ANITA-tuffs}. 

For this study we apply a simplified model of the ANITA trigger. Given a time-domain electric field peak $E_\mathrm{peak}$, we approximate the peak voltage $V_\mathrm{peak}$ at the detector using
\begin{equation}
V_\mathrm{peak}=E_\mathrm{peak}\frac{c}{f_c}\sqrt{\frac{R_L}{Z_0}\frac{D}{4\pi}},
\end{equation}
where $R_L=50\,\Omega$ is the load impedance of the ANITA receiver, 
$Z_0=377\,\Omega$ is the impedance of free space, and $D=10$~dBi is the peak directivity of the ANITA horn antennas. We assume a central frequency, $f_c$, of 300~MHz for the conversion. 

We estimate the detector threshold based on the weakest event in the population of cosmic-ray air showers detected in ANITA-I (reported in \cite{Hoover_dissertation}) and ANITA-III. The smallest peak electric field in ANITA-I (ANITA-III) reported was $E_\mathrm{peak}=$~446~ (284)~$\mu$V/m. This corresponds to a threshold voltage of $V_\mathrm{peak}=$~143~(91)~$\mu$V. The improvements to the ANITA-III instrument result in a factor of $\sim$2 decrease in the estimated trigger threshold compared to ANITA-I. The trigger is approximated by taking $p_\mathrm{trig}$ to be unity if the electric field is above this threshold and zero if it is below.


\subsection{Monte Carlo simulations}
The acceptance in Equation~\ref{eq:acceptance} is evaluated via Monte Carlo integration. The total region of
integration is given by the detector horizon, characterized by 
$\cos\theta_{E,\mathrm{horz}}=(1+h/R_E)^{-1}$ (see Section 3 and Figure~\ref{fig:geom}). The maximal aperture for the region of integration is given by \cite{Motloch-Privitera}
\begin{equation}
A_0 \simeq 2\pi^2 \frac{R_E h}{1+h/R_E} .
\end{equation}

Given the geometry of the detector, in the simulation we sample the set of parameters \{$\theta_E$, $\theta_{\nu_\tau}$, $\phi_{\nu_\tau}$,  $\mathcal{E}_\tau$, $s_\mathrm{decay}$\}.
The location of the exit point of the particle on the surface of the Earth is obtained from sampling the polar angle with respect to the position of the detector ($\theta_E$) from a cosine distribution in the interval ${[(1+h/R_E)^{-1},1]}$, according to the integral in Eq.~(\ref{eq:acceptance}). Since the integrand in Eq.~(\ref{eq:acceptance}) is azimuthally symmetric around the axis of the detector, $\phi_E$ need not be sampled. The particle trajectory vector $\hat{r}_{\nu_\tau}$ is obtained from sampling its polar coordinate parameters $\theta_{\nu_\tau}$ and $\phi_{\nu_\tau}$. Due to the dot product of $\hat{r}_{\nu_\tau}\cdot \hat{n}_E$ in Eq.~(\ref{eq:acceptance}), the angle $\theta_{\nu_\tau}$ is sampled
according to a cosine-squared distribution in the interval $[0,1]$, since we consider only
exiting trajectories in the field of view of the detector. The azimuthal angle $\phi_{\nu_\tau}$ is uniformly sampled in
the interval $[0,2\pi]$. The exit angle $\theta_\mathrm{exit}$ is obtained from $\theta_E$ and $\hat{r}_{\nu_\tau}$. The $\tau$ lepton energy $\mathcal{E}_\tau$ is sampled from a distribution obtained with a separate tau neutrino propagation simulation (see Section~\ref{sec:tau_exit_prob}) for the corresponding exit angle. The decay distance in the atmosphere is obtained from sampling $s_\mathrm{decay}$ from the probability distribution $p_\mathrm{decay}(s_\mathrm{decay}|\mathcal{E}_\tau)$. As mentioned in the beginning of Section 3, we assume $\mathcal{E}_\mathrm{EAS}=\mathcal{E}_\tau$ and propagate the corresponding electric field to the location of the detector.
The probability that an event is detected $p_\mathrm{det}(\mathcal{E}_{\tau,k}, \hat{r}_{\nu_\tau,k}, \theta_{E,k})$ includes the sampling of $p_\mathrm{decay}$, $p_\mathrm{EAS}$, $p_\mathrm{pk}$ and $p_\mathrm{trig}$. In this simulation $p_\mathrm{trig}$ is 1 if the event is above threshold and 0 if it is below.
The numerical estimate of the acceptance is
\begin{equation}
\begin{split}
\langle A\Omega\rangle_{\nu_\tau}(\mathcal{E}_{\nu_\tau})
= 
\frac{A_0}{N} 
\ \sum_{k=1}^{N} &
\ P_\mathrm{exit}(\mathcal{E}_{\nu_\tau}, \theta_{\mbox{\scriptsize exit},k}) 
\\ &
\  \times p_\mathrm{det}(\mathcal{E}_{\tau,k}, \hat{r}_{\nu_\tau,k}, \theta_{E,k}),
\end{split}
\end{equation}
where the index $k$ labels each of the $N$ simulated events. 
The marginalized $\tau$-lepton exit probability 
$P_\mathrm{exit}(\mathcal{E}_{\nu_\tau}, \theta_\mathrm{exit})$, defined in Eq.~(\ref{eq:exitprob}),
accounts for the fact that we sampled an exiting tau lepton probability with
energy $\mathcal{E}_\tau$ including the tau neutrinos that do not
result in a tau lepton exiting the surface of the Earth.\\

\begin{figure*}[htbp]
  \centering
\includegraphics[width=0.45\textwidth]{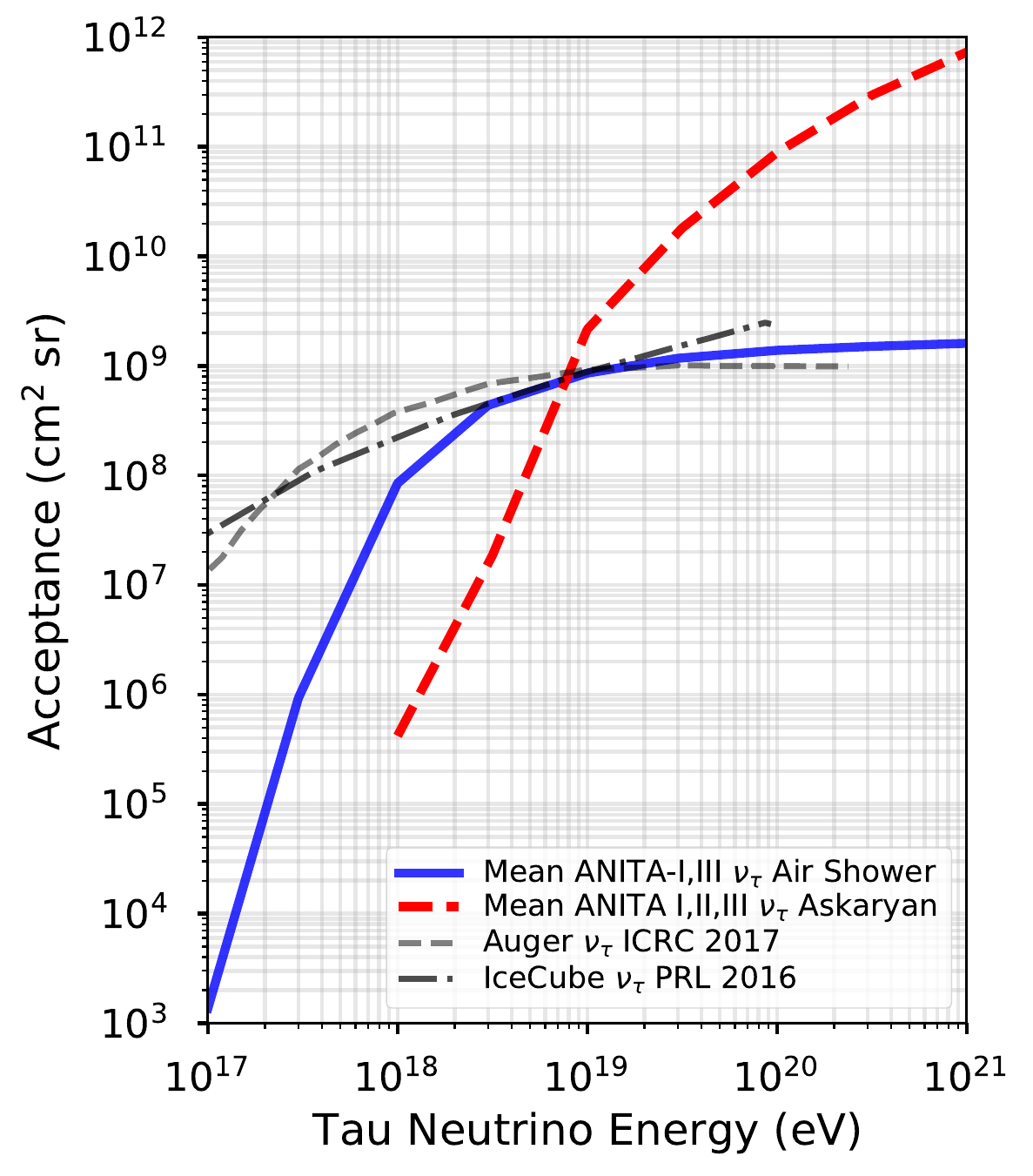}
\includegraphics[width=0.45\textwidth]{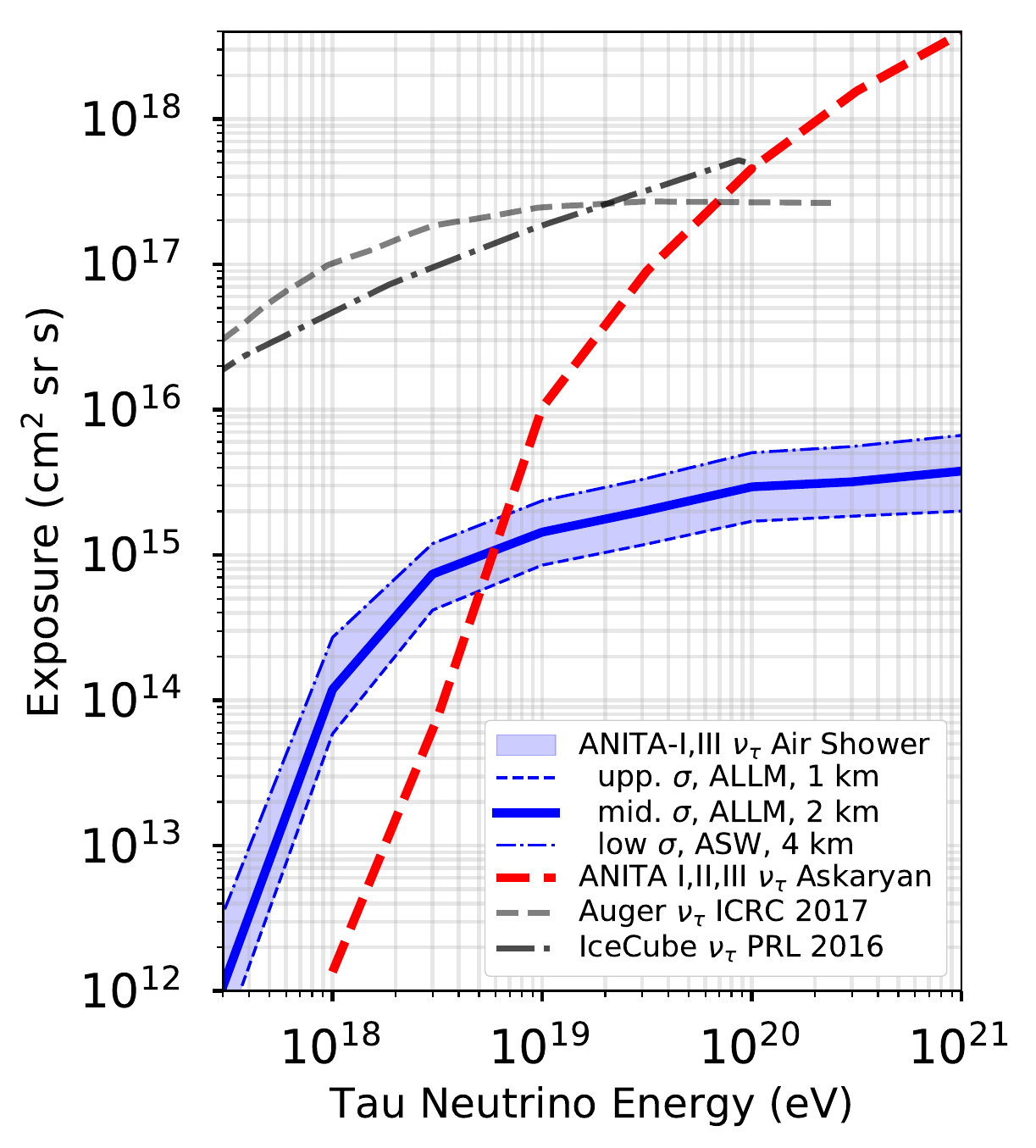}
   \caption{\raggedright
   Monte Carlo-derived upper bound estimates of the ANITA acceptance and exposure to tau neutrinos.
   Left: The mean acceptance of ANITA-I and ANITA-III to $\tau$-lepton air showers of $\nu_\tau$ origin (blue solid line) assuming standard values of cross section and energy loss models and 2.0 km ice thickness. These are compared to Auger (dashed grey line), IceCube (dot-dashed darker grey line), and the ANITA Askaryan search for in-ice showers from $\nu_{\tau}$'s (red dashed line). 
Right: Upper bounds on the ANITA exposure (blue solid line). The blue-shaded band includes the range of variations due to assumptions on the ice thickness (1-4 km), neutrino cross section, and $\tau$ energy loss models. The minimum exposure (dashed blue line) assumes a high cross section, ALLM~\cite{ALLM_97} energy loss model, and 1~km ice thickness, while the maximum exposure (dot-dashed blue) assumes a low cross section, ASW~\cite{ASW_2005} energy loss model, and 4~km ice thickness. The exposure to standard values for the cross section and energy loss model (ALLM) and the average ice thickness of 2~km is shown with a solid blue line. 
For comparison, we include the ANITA Askaryan exposure to $\nu_{\tau}$'s (red dashed line) \cite{ANITA-3-askaryan}, Auger 2017 (dashed grey) exposure to Earth-skimming tau neutrinos \cite{Auger_2017}, and IceCube 2016 (dot-dashed darker grey) exposure to tau neutrinos \cite{IceCube_2016}. Note that the solid blue line is the only fair comparison for the standard neutrino cross section and energy loss models; otherwise the Auger, IceCube, and ANITA Askaryan exposure curves would also have to be modified.}
\label{fig:exposures}
\end{figure*}

\section{Results}

\subsection{Upper bound on exposure}
The resulting upper bounds on the ANITA acceptance and exposure to $\tau$-lepton air showers of $\nu_\tau$ origin are shown in Figure~\ref{fig:exposures} (labeled Air Shower). Loss of sensitivity due to the effects of phi-masking and deadtime are included in the exposure estimates, but not in the acceptance estimate. The $\tau$-lepton air shower acceptance upper bound curve on the left panel of Figure~\ref{fig:exposures} is obtained from simulations using the ANITA-I threshold and the ANITA-III threshold and taking the arithmetic mean.  
At energies $\mathcal{E}_{\nu_\tau}>3\times10^{18}$~eV this upper bound estimate  is 
 comparable to the $\nu_\tau$ acceptances of IceCube and Auger. With decreasing energy $\mathcal{E}_{\nu_\tau}<3\times10^{18}$~eV, the ANITA acceptance falls off quickly making ANITA orders of magnitude less sensitive. The average ANITA acceptance curve for $\nu_\tau$ interacting in the ice sheet and producing a coherent radio impulse exiting the ice (labeled Askaryan) is also shown for comparison~\cite{ANITA-3-askaryan}. At energies $\mathcal{E}_{\nu_\tau}>10^{19}$~eV the acceptance of the Askaryan channel is significantly larger but decreases more steeply with decreasing neutrino energy than the air shower channel.

The curves on the right panel of Figure~\ref{fig:exposures} show that the ANITA $\tau$-lepton air shower channel for $\nu_\tau$ has a substantially lower exposure compared to IceCube and Auger. This is primarily due to the fact that IceCube and Auger have run continuously for many ($\sim 10$) years. The blue band for the ANITA $\tau$-lepton air shower channel brackets the range of curves obtained from ice shell thicknesses between 1 and 4~km as well as the range of $\nu_\tau$ cross sections and $\tau$ energy loss models considered in this work (see \cite{Alvarez-Muniz_2018} for more details). The ANITA $\tau$-lepton air shower exposure is at least a factor of 40 smaller than Auger or IceCube at high energies and more than four orders of magnitude smaller at relevant energies $\sim 3\times 10^{17}$~eV.

\begin{figure*}[htbp]
  \centering
\includegraphics[width=\textwidth]{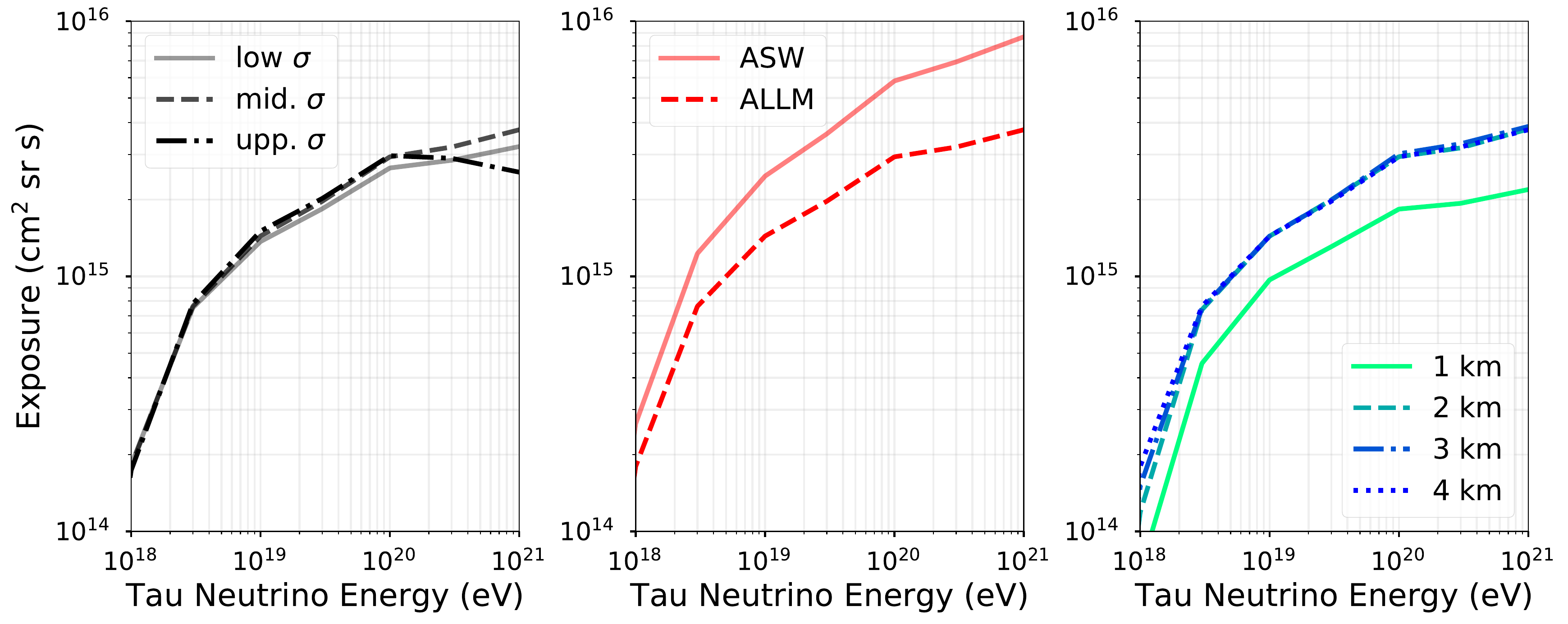}

   \caption{\raggedright 
   The upper bounds of the ANITA exposure to upward-going tau neutrino-induced air showers assuming (left) variations on the cross section for 4-km thick ice and the ALLM energy loss model, (middle) variations on the energy loss model for 4-km thick ice and the mid-range cross section, and (right) variations on the ice thickness for the mid-range cross section and ALLM energy loss model. }
   \label{fig:exposure_var}
\end{figure*}

In Figure~\ref{fig:exposure_var}, we show the dependence of the exposure of the ANITA $\tau$-lepton air shower channel on neutrino interaction cross section, $\tau$-lepton energy-loss models, and ice thickness. In the left panel, we show that the upper and lower uncertainties on the cross section in~\cite{Connolly_2011} have a small effect on the exposure at neutrino energies $<10^{20}$~eV. At energies $\mathcal{E}_{\nu_\tau}\simeq10^{21}$~eV, the exposure varies by $\sim 70\%$. 
As discussed in~\cite{Alvarez-Muniz_2018}, increasing (decreasing) the cross-section increases (decreases) $p_\mathrm{exit}$ for emergence angles below the value corresponding to the trajectory being tangential to rock beneath the ice layer while for emergence angles above this value $p_\mathrm{exit}$ decreases (increases). The standard (mid.) value of the cross-section happens to maximize the probability of detection integrated over all emergence angles at $\mathcal{E}_{\nu_\tau}\simeq10^{21}$~eV.  

In the middle panel of Figure~\ref{fig:exposure_var} we compare the exposures obtained with the ALLM~\cite{ALLM_97} and ASW~\cite{ASW_2005} $\tau$ energy loss models. The ASW model, with a lower $\tau$-lepton energy loss, results in a larger acceptance. This is the largest contribution to the uncertainty within the Standard Model which is of order a factor of $\sim2$ for $\mathcal{E}_{\nu_\tau}\simeq10^{19}$~eV. A reduced energy loss increases the $\tau$ decay range (energy loss and decay combined), thus enabling a larger interaction volume near the surface of the Earth to contribute to exiting $\tau$ leptons~\cite{Alvarez-Muniz_2018}. 

Finally, in the right panel of Figure~\ref{fig:exposure_var} we display the dependence of the exposure on the thickness of the ice above sea level. As the ice thickness increases from 1~km to 2~km in addition to the Earth's radius, the exposure increases by a factor of $\sim$~2 for energies above $5\times10^{18}$~eV. Since thicker ice increases the altitude above sea level that the tau emerges into, increasing the ice thickness above 2~km does not further increase the exposure. This is due to the competing effects of an increased $P_\mathrm{exit}$ with thicker ice~\cite{Alvarez-Muniz_2018} versus a weaker air shower electric field strength due to the thinner atmosphere at higher altitude above sea level.
For neutrino energies below $10^{19}$~eV, the difference between a 1~km and 4~km ice shell is small while at higher energies the effect increases  but remains smaller than a factor of two.

\subsection{Differential acceptance vs. emergence angle}
\begin{figure*}[!ht]
\centering
\includegraphics[width=0.45\textwidth]{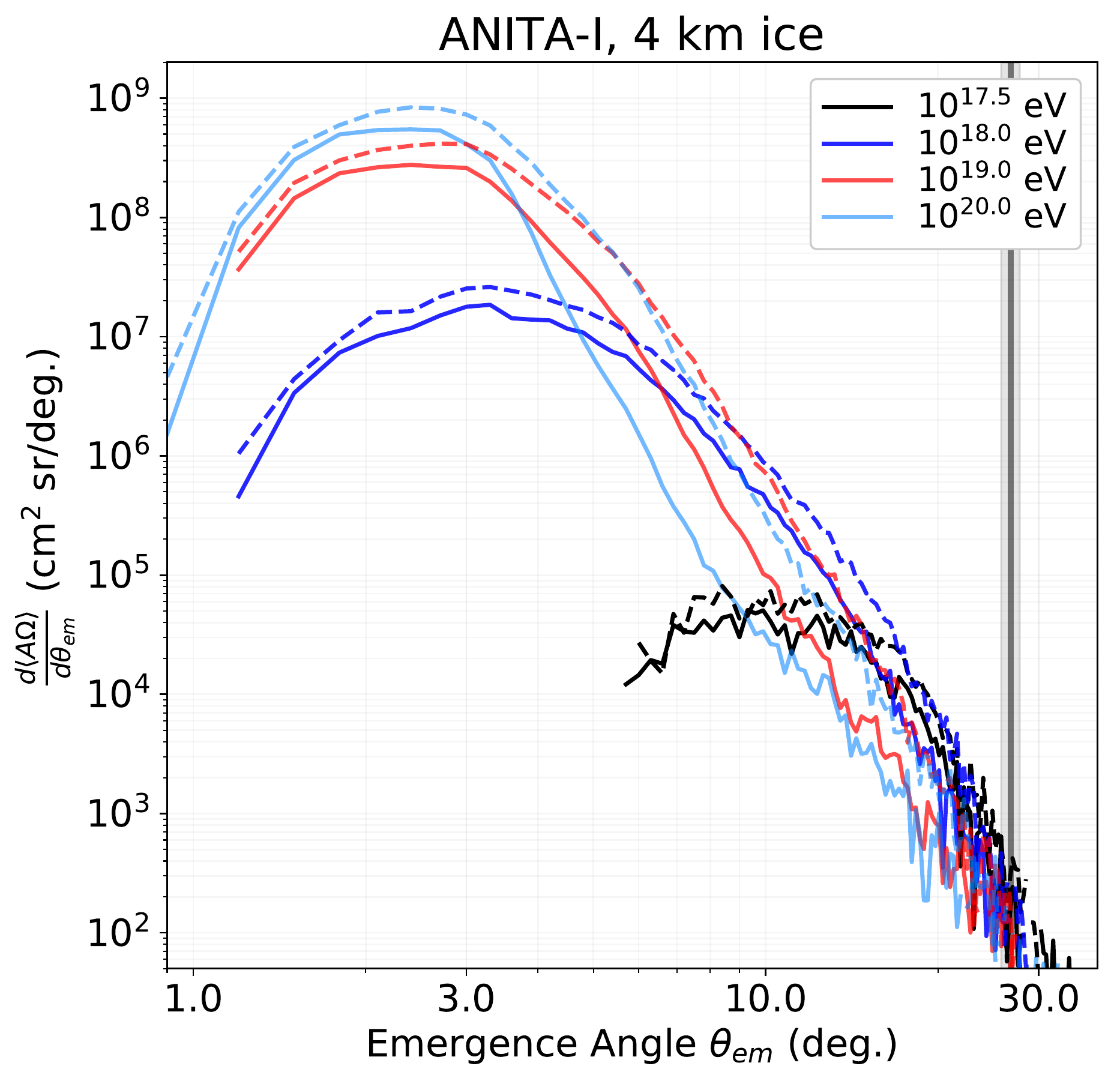}
\includegraphics[width=0.45\textwidth]{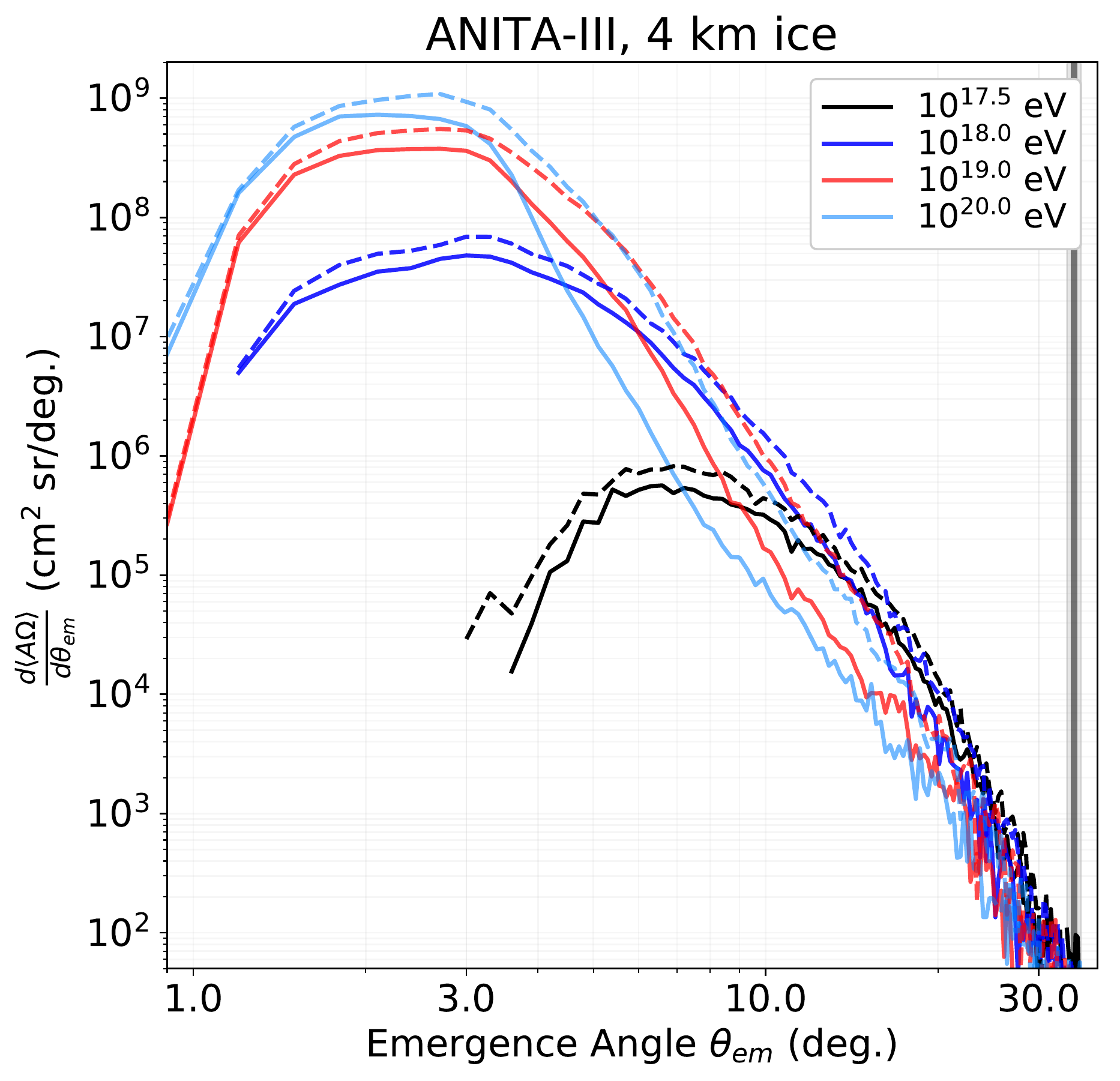}
\caption{\raggedright 
ANITA-I  (left) and ANITA-III (right) differential acceptance vs. emergence angle for 4~km ice thickness for various energies. The mid-range Standard Model cross section with the ALLM energy loss model are shown as solid lines and the low-range Standard Model cross section with ASW energy loss model are shown as dashed lines. The reconstructed emergence angles for the ANITA events and their uncertainties projected to the ice are shown in the vertical band with a line.}
\label{fig:diffacceptance}
\end{figure*}

To further compare the simulations to the observed events, in Figure~\ref{fig:diffacceptance} we show ANITA's differential acceptance to an isotropic tau neutrino flux as a function of emergence angle. The most optimistic case of an ASW energy loss model and the lowest Standard Model cross section (dashed lines) results in a broader differential acceptance that extends to wider emergence angles when compared with the results from a mid-range Standard Model cross section and ALLM energy loss model (solid lines). The lower trigger threshold of ANITA-III increases the differential acceptance at all energies to higher emergence angles when compared to ANITA-I. At the lowest energies ($\leq 10^{18}$~eV), the lower trigger threshold increases the total acceptance by factors of 5-10 and shifts the peak in the differential acceptance to lower emergence angles.

The emergence angles for the ANITA-I and ANITA-III events, shown in Figure~\ref{fig:diffacceptance} as a vertical line, are in the tails of the estimated differential acceptance for both ANITA-I and ANITA-III. 
At neutrino energies $\geq10^{18}$~eV, the differential acceptance is $\sim5$ (ANITA-I) and $\sim6$ (ANITA-III) orders of magnitude higher at emergence angles between $2^{\circ}-5^{\circ}$ than at $25^{\circ}$ or larger (where the ANITA events lie). This means that if the observed events were due to an isotropic flux, the neutrino energy has to be $<10^{18}$~eV. Otherwise, more events would be expected at low emergence angles.

For a $\nu_\tau$ energy of $\sim10^{17.5}$~eV, the ANITA-I event is $\sim$100 times more likely to emerge at $\sim10^{\circ}$ compared to the observed emergence angle of 25.4$^{\circ}$.  
For ANITA-III, the differential acceptance at $10^{17.5}$~eV is a factor of $>$1000 higher at 10$^{\circ}$ than at the observed emergence angle of $34.6^{\circ}$. 

For the hypothesis of a Standard Model $\tau$-lepton of $\nu_\tau$ origin of ANITA anomalous events to be consistent with the data, substantially more events would be expected at low emergence angles. Further suppression of the cross section, beyond the Standard Model (see for example~\cite{Cornet_2001, Jain_2002,Reynoso_2013}), would further shift the peak of the distribution to larger emergence angles. This will be the subject of a future study. It is worth noting that the upper bound approach taken here tends to overestimate the acceptance and increasing the fidelity of the detector model will reduce the sensitivity, particularly at the high emergence angles where the ANITA antenna beam pattern tends to lose gain.



\section{Discussion \& Conclusions}


In this study we have placed an upper bound on the ANITA-I and ANITA-III exposure to $\tau$-lepton air showers of $\nu_\tau$ origin and compared it to IceCube, Auger, and the ANITA in-ice shower channel. The ZHAireS simulation code was adapted to produce upgoing air showers from $\tau$ lepton decays in the atmosphere, which enabled a Monte Carlo upper bound estimate of the exposure. The code, which could be used for other $\tau$-lepton detector simulations such as~\cite{GRAND, TAROGE}, is available upon request to the authors. The possible radio emission profiles for the specific ANITA-I and ANITA-III events have been presented and a lower limit on the energy of the air showers are estimated in both cases to be above 2.5$\times10^{17}$~eV.

The main conclusion is that the observation of $\tau$-lepton events from a diffuse neutrino flux by the ANITA flights is inconsistent with the limits placed by IceCube and Auger with Standard Model parameters by several orders of magnitude. Although the acceptance of ANITA is smaller than but comparable to IceCube and Auger, the significantly higher duty cycle of these observatories makes their exposure more than two orders of magnitude higher than ANITA at neutrino energies above $10^{19}$~eV and significantly more at energies below that. The constraints include a characterization of the dependence on ice thickness, neutrino-nucleon cross section uncertainties, and $\tau$-lepton energy loss models, all within the Standard Model. Although these effects can modify the exposure upper bounds by a factor of 2 to 5, depending on the energy, it is not enough to address the strong tension with the IceCube and Auger $\nu_\tau$ flux bounds.  

The $\nu_\tau$ cross section and the $\tau$-lepton energy loss models used in this study are by no means exhaustive. Significant dependence of these models on the exposure has been shown. It is possible that with more aggressive suppression of the cross section compared to the Standard Model the discrepancy with IceCube and Auger 
might be reduced. However, for such a study to be conclusive, it would require estimates of the IceCube and Auger exposure with the same modified interaction models for fair comparison. 

Despite ANITA's exposure in this $\nu_\tau$ air shower channel being smaller than IceCube and Auger, the acceptance is comparable to those observatories at energies $>10^{18}$~eV. This is indicative that ANITA may be highly sensitive to point source fluxes and transients. This will be explored in detail in a follow-up paper.

The Standard Model $\tau$-lepton of a diffuse $\nu_\tau$ flux origin hypothesis is not self consistent within ANITA observations. The expected emergence angle from this model is significantly smaller than the observed emergence angles. It is possible that this discrepancy could be reduced by a more aggressive suppression of the neutrino-nucleon cross section, as has been suggested in some beyond Standard Model scenarios~\cite{Cornet_2001, Jain_2002,Reynoso_2013}. The effect will reduce the $\tau$-lepton exit probability at lower emergence angles in favor of higher emergence angles. 
Other possibilities that could resolve this discrepancy include sterile neutrinos~\cite{Huang_2018}, the decay in Earth of a quasi-stable dark matter particle~\cite{Anchordoqui_2018}, and supersymmetric sphaleron transitions~\cite{Anchordoqui_2019}.
This will be treated in a future study.

ANITA-IV had a longer flight than ANITA-I and ANITA-III and the analysis of its data is currently underway. The continued detection of radio impulses consistent with up-going air showers will motivate more detailed studies of the origin of these events.

\noindent{\it Acknowledgements:} 
Part of this work was carried out at the Jet Propulsion Laboratory, California Institute of Technology, under a contract with the National Aeronautics and Space Administration.
We thank NASA for their generous support of ANITA,
the Columbia Scientific Balloon Facility for their excellent
field support, and the National Science Foundation for their
Antarctic operations support. This work was also supported
by the U.S. Department of Energy, High Energy Physics
Division.
S.~A.~W. thanks the National Science Foundation for support through award \#1752922.
J. A-M and E.Z. thank 
Ministerio de Econom\'\i a, Industria y Competitividad (FPA 2017-85114-P),
Xunta de Galicia (ED431C 2017/07),
Feder Funds, 
RENATA Red Nacional Tem\'atica de Astropart\'\i culas (FPA 2015-68783-REDT) and
Mar\'\i a de Maeztu Unit of Excellence (MDM-2016-0692).
W.C. thanks grant \#2015/15735-1, S\~ao Paulo Research Foundation (FAPESP).
We thank N. Armesto and G. Parente for fruitful discussions on the neutrino cross-section and $\tau$ lepton energy-loss models.
\noindent\textsf{\copyright} 2019. All rights reserved.



\begin{thebibliography}{99}

\bibitem{ANITA_up}
P. Gorham {\it et al.} [ANITA Collaboration], Phys. Rev. Lett. {\bf 117}, 071101 (2016).

\bibitem{ANITA3_tau}
P. Gorham {\it et al.} [ANITA Collaboration], Phys. Rev. Lett. {\bf 121}, 161102 (2018).

\bibitem{ANITA_CR}
S. Hoover {\it et al.} [ANITA Collaboration], Phys. Rev. Lett. {\bf 105}, 151101 (2010).

\bibitem{ZHAireS}
J. Alvarez-Mu\~niz {\it et al.}, Astropart. Phys. {\bf 35}, 325-341 (2012).

\bibitem{ZHAireS_UHF}
J. Alvarez-Mu\~niz {\it et al.}, Phys. Rev. D {\bf 86}, 123007 (2012).

\bibitem{ZHAireS_superposition}
J. Alvarez-Mu\~niz {\it et al.}, Astroparticle Physics, {\bf 59}, 29 (2014).

\bibitem{ZHAireS_reflected}
J.~Alvarez-Mu\~niz, W.R.~Carvalho Jr., D. Garc\'ia-Fern\'andez, H. Schoorlemmer, and E.~Zas, Astropart. Phys. {\bf 66}, 31-38 (2015).

\bibitem{Belov_2016}
K.~Belov {\it et al.}, [The T-510 Collaboration], Phys. Rev. Lett. {\bf 116}, 141103 (2016).

\bibitem{ANITA_spectrum}
H. Schoorlemmer {\it et al.} [ANITA Collaboration], Astropart. Phys. {\bf 86}, 32-43 (2016).

\bibitem{Auger_spectrum}
The Pierre Auger Collaboration, Proceedings of the 33rd International Cosmic Ray Conference, Rio de Janeiro, 2013, arXiv:1307.5059

\bibitem{TA_spectrum}
T. Abu-Zayyad {\it et al.} [Telescope Array Collaboration], Astropart. Phys. {\bf 61}, 93101 (2015).

\bibitem{Auger_2015}
A. Aab {\it et al.}, [The Pierre Auger Collaboration],
Phys. Rev. D {\bf 91}, 092008 (2015).

\bibitem{Auger_2017}
E. Zas for the Pierre Auger Collaboration in
Proceedings of the $35^\mathrm{th}$ International Cosmic Ray Conference, 
PoS(ICRC2017)972.

\bibitem{IceCube_2016}
M.~G.~Aartsen {\it et al.}, 
Phys. Rev. Lett {\bf 117}, 241101 (2016).

\bibitem{Alvarez-Muniz_2018}
J.~Alvarez-Mu\~niz, W.~R.~Carvalho Jr., K. Payet, A. Romero-Wolf, H. Schoorlemmer, and E.~Zas, Phys. Rev. D {\bf 97}, 023021 (2018).

\bibitem{tauola} S. Jadach {\it et al.}, 
Comput. Phys. Commun. {\bf 76}, 361 (1993).

\bibitem{ZHS92}
E.~Zas, F.~Halzen, and T.~S.~Stanev, Phys. Rev. D, {\bf 45}, 362 (1992).

\bibitem{Garcia-Fernandez2013}
D.~Garc\'\i a-Fern\'andez, J.~Alvarez-Mu\~niz, W.~R.~Carvalho Jr., A. Romero-Wolf, and E.~Zas, Phys. Rev. D {\bf 87}, 023003 (2013).

\bibitem{ANITA_instrument_paper}
P. Gorham {\it et al.} [ANITA Collaboration], Astropart. Phys. {\bf 32}, 10-41 (2009).

\bibitem{ANITA-3-askaryan}
P. Gorham {\it et al.} [ANITA Collaboration], 
Phys. Rev. D {\bf 98}, 022001 (2018).

\bibitem{ANITA-tuffs}
P.~Allison, O.~Banerjee, J.~Beatty, A.~Connolly {\it et al.} [ANITA Collaboration], 
NIM-A {\bf 894}, 47-56 (2018).

\bibitem{Hoover_dissertation}
S. Hoover, Ph.D. thesis, UCLA (2010).

\bibitem{Motloch-Privitera}
P. Motloch, N. Hollon, and P. Privitera,
Astropart. Phys. {\bf 54}, 40 (2014).


\bibitem{Connolly_2011} 
A. Connolly, R. S. Thorne, D. Waters, 
Phys. Rev. D {\bf 83}, 113009 (2011).

\bibitem{ALLM_97} 
H. Abramowicz and A. Levy, 
arXiv: hep-ph/9712415.

\bibitem{ASW_2005} 
N. Armesto, C. Salgado, and U.~A.~Wiedemann, 
Phys. Rev. Lett. {\bf 94}, 022002 (2005).

\bibitem{Cornet_2001}
F. Cornet, J.~I. Illana, M. Masip,
Phys. Rev. Lett. {\bf 86}, 4235 (2001).

\bibitem{Jain_2002}
A. Jain, P. Jain, D. W. McKay, J. P. Ralston,
Int. Jour. of Mod. Phys. A {\bf 17}, 533 (2002)

\bibitem{Reynoso_2013} 
M.~M.~Reynoso, O.~A.~Sampayo, 
J. Phys. G: Nucl. and Part. Phys. {\bf 83}, 113009 (2011)

\bibitem{GRAND} 
J. Alvarez-Mu\~niz {\it et al.} [GRAND Collaboration], 
arXiv:1810.09994 (2018).

\bibitem{TAROGE}
J. Nam,
Proceedings of the $34^\mathrm{th}$ International Cosmic Ray Conference,
PoS(ICRC2015)663. 

\bibitem{Huang_2018}
G-Y.~Huang,
Phys. Rev. D {\bf 98}, 043019, (2018)

\bibitem{Anchordoqui_2018}
 L.~A.~Anchordoqui, V.~Barger, J.~G.~Learned, D.~Marfatia and T.~J.~Weiler,
LHEP {\bf 1}, 13, (2018)

\bibitem{Anchordoqui_2019}
 L.~A.~Anchordoqui and I.~Antoniadis,
arXiv:1812.01520, accepted for publication in Phys. Lett. B,  (2019)


\end{thebibliography}
\end{document}